# H I, H II, AND R-BAND OBSERVATIONS OF A GALACTIC MERGER SEQUENCE

J. E. Hibbard[1]

*Institute for Astronomy*
*2680 Woodlawn Drive, Honolulu, HI 96822*
*hibbard@gmc.ifa.hawaii.edu*

J. H. van Gorkom

*Astronomy Department, Columbia University*
*538 W 120 Street, New York, NY 10027*
*jvangork@fidelio.phys.columbia.edu*

## Abstract

We present high-quality aperture synthesis observations of the neutral hydrogen distribution in a sample of five galactic systems believed to be involved in progressive stages of merging: Arp 295, NGC 4676, NGC 520, NGC 3921, and NGC 7252. This is the first time that the atomic hydrogen in such a broad range of disk mergers has been imaged. These data are supplemented by wide-field images taken through a narrow band Hα filter, and by deep ($\mu_R \gtrsim 26.5$ mag arcsec$^{-2}$) R-band surface photometry.

We identify several trends along the merging sequence. In the early stages, large amounts of H I still exist within the galactic disks and star formation is widespread. The ionized gas emission often takes on the appearance of plumes and arcs emanating from the nuclear regions, which are presumably the sites of interaction-induced starbursts. In the final stages there is little if any H I within the remnant bodies, and tidal material is seen moving inward. We conclude that as the merger rearranges the light profiles of the progenitor disk galaxies into $r^{1/4}$ profiles, it leads to an efficient conversion of the atomic gas into other forms within the main bodies of the merger remnants. This suggests that these remnants will evolve into elliptical galaxies in their atomic gas contents as well as their photometric properties. However the observations of NGC 520 reveal an extensive rotating gaseous disk, suggesting that perhaps some mergers will not destroy the atomic gas disks of the progenitors.

Although the observed tidal features contribute only of order 10% of a system's optical luminosity, they contain the majority of the neutral hydrogen. The morphological similarity between the gaseous and stellar tails and the smooth gas kinematics confirm that gravity plays the dominant role in producing them. There are, however, some striking differences between the faint gas and light distributions. H I mapping often reveals gaseous extension not at all apparent optically, and tidal features of different optical morphologies have different gas characteristics: the edge-brightened tails are gas-rich, while the featureless plumes and halos are gas-poor. Some of these features may be explained by the different velocity dispersions of the gas and stars, some by different gas contents in the progenitors, and some remain unexplained. Overall, large quantities of both gas and starlight ($M_{\rm HI} > 10^9\,h^{-2}\,M_\odot$, $L_R > 10^9\,h^{-2}\,L_\odot$) are seen at large radii ($r > 50\,h^{-1}$ kpc). Since this material evolves on very long time scales, it may leave observable signatures for many Gyr.

*Subject headings*: galaxies: individual (Arp 295, NGC 4676, NGC 520, NGC 3921, NGC 7252) — galaxies:interactions — galaxies: mergers — galaxies: interstellar matter — galaxies: H I — galaxies: CCD photometry—galaxies: ionized gas

---

[1] Hubble Fellow

# 1. Introduction

Over two decades ago, Toomre (1970) and Toomre & Toomre (1972; hereafter referred to as TT) presented simple numerical models which demonstrated that gravitational interactions between galaxies could generate tidal features, such as the tails and bridges resplendent in Arp's (1966) Atlas. Both Toomre & Toomre (1972) and Toomre (1977) proposed that strong collisions between disk galaxies would lead to orbital decay and to eventual merging. In 1977, Toomre presented 11 peculiar galaxies that span a proposed evolutionary sequence of galactic merging (the "Toomre Sequence"). He explained how the end-product might be an elliptical galaxy, an idea which has come to be known as the merger hypothesis.

This hypothesis has been vindicated through careful observational studies of Toomre-like objects, showing them to have the optical characteristics (Schweizer 1978, 1980, 1981, 1982, 1995) and kinematics (Stockton 1974a,b; Schweizer 1978; Stockton & Bertola 1980) expected from a merger origin. On the numerical front, increasingly sophisticated theoretical calculations have been made possible by the rapid advances in computing hardware and numerical techniques (*cf.* Barnes & Hernquist 1992b and references therein). In particular, it has been demonstrated numerically that two model disk galaxies can indeed suffer strong enough dynamical friction and violent relaxation to merge rapidly and redistribute their "stars" into a relaxed profile typical of elliptical galaxies (Farouki & Shapiro 1982, Barnes 1988). The inclusion of gas-dynamical effects suggests that the high central phase space densities of ellipticals (Ostriker 1980) may be obtained through the dissipation and subsequent star formation of the gaseous component (TT, Carlberg 1986, Negroponte & White 1983).

Much observational work has emphasized the effect of merging on the ample interstellar medium (ISM) of the spiral progenitors. The colors and spectra of galaxies with tidal distortions provide evidence of tidally induced starburst (see review by Kennicutt 1990). $^{12}$CO(1→0) mapping has shown that the most luminous infrared galaxies, thought to be gas rich systems in the throes of merging (Lonsdale *et al.* 1984, Joseph & Wright 1985, Schweizer 1986, Sanders *et al.* 1988b), have central gas contents and densities far higher than their spiral progenitors (see review by Scoville *et al.* 1993), implying significant dissipation of the gaseous component (Kormendy & Sanders 1992). Further, the association of radio, Seyfert, and QSO activity with the presence of tidal distortions, companions, and/or red-shifted H I absorption suggests interaction induced gaseous infall as a likely meal for lurking central monsters (see reviews by Heckman 1990, Stockton 1990).

However, observational work systematically documenting the effects of merging upon the atomic gas of the spiral progenitors has lagged far behind both the numerical work and the observational studies of mergers at other wavelengths. In an attempt to rectify this situation, we have obtained high quality H I data on five present or former disk-disk systems in different stages of merging. Four of these systems have been taken from the Toomre Sequence.

The goal of this study is to investigate the fate of the gas in mergers by studying the distribution of the cold neutral atomic gas, the warm ionized gas and the starlight. To this end, radio synthesis observations are used to obtain good spatial resolution, high sensitivity to extended gas emission, and line-of-sight velocity information. Broad-band $R$ images are used to delineate the underlying stellar distribution and to uncover any faint optical tidal extensions. We choose to image the faint light in the $R$-band so as to be less biased towards younger stellar components than in $B$ or $V$. Narrow-band H$\alpha$ observations reveal regions of current star formation.

These data allow us to investigate the physical impact of strong interactions and merging upon the structure and state of the ISM. The distribution of the faint optical light and gas reflects the gross effect of the interaction on the progenitors, and provides clues to their identity. A comparison of the starlight with the neutral hydrogen may identify where and when hydrodynamical effects are important, and the distribution of the ionized gas identifies regions of high energy input into the ISM. A comparison of both the neutral and ionized hydrogen, when combined with existing data on the molecular gas content and distribution, helps identify regions of phase transitions in the gas and the location of future star-forming material. These techniques are fully demonstrated in the study of the NGC 7252 merger remnant (Hibbard *et al.* 1994, hereafter referred to as HGvGS).

The complete velocity mapping of the tidal features provided by the neutral hydrogen observations is uniquely suited for studies of disk-disk encounters. This is because (1) the tidal features are drawn primarily from the outer regions (TT) which, for disk galaxies, are rich in atomic gas (Wevers *et al.* 1986); and (2) this outer gas has not undergone more than 1–2 rotations since the encounter started and therefore retains spatial and kinematic signatures of the encounter geometry (TT; Barnes & Hernquist 1992b). The tidal features therefore bear the imprint of the encounter history, and their morphology allows us to constrain their space velocities to help elucidate the past and future evolution of such systems. This technique has been successfully applied to the NGC 7252 data (Hibbard & Mihos 1995), and similar studies investigating the kinematics of each member of the chosen sequence will be published elsewhere.

The present paper is structured as follows: §2 describes the sequence chosen for study; §3 describes the observations and data reduction techniques; §4 presents the results in terms of figures and tables; §5 describes the trends observed along the sequence; §6 discusses these data, with the goal of understanding the fate of the gas along the sequence; and §7 summarizes the main results. In the appendix, we present the H I channel maps and a more detailed description of each system.

## 2. The Evolutionary Sequence

The systems in this study are primarily chosen from the "Toomre Sequence" (Toomre 1977; see also TT) of the 11 foremost examples of ongoing disk-disk mergers from the NGC[1]. Each member of this sequence was selected for its pair of well defined tidal extensions, implicating two disk galaxies as the progenitors. The sequence forms the backbone of the merger scenario, which has since been widely supported by both observational and numerical work (for reviews see Schweizer 1990; Barnes & Hernquist 1992b). In this scenario, two galaxies lose their mutual orbital energy and angular momentum and coalesce into a single object, the merger remnant. It is in the context of this scenario that the sequence is evolutionary, with the systems arranged according to the putative time remaining until the final merger. This time is judged by the degree of coalescence of the progenitors main bodies, assuming that they were normal disk galaxies in the pre-encounter state.

We have chosen four representative systems from the Toomre sequence: NGC 4676, NGC 520, NGC 3921, and NGC 7252. To these we add the bridge-tail system Arp 295, in order to map the gas distribution in a system in which the progenitors are well separated but still likely to merge. It is an obvious choice for inclusion here since it is one of the four systems modeled by the Toomres in the original bridge-and-tail building paper (TT). The five systems span the full range of the proposed merging sequence, and are shown as negative greyscale images in Figure 1(a–e), reproduced from the $R$-band optical data described later in this paper. A bar representing 20 $h^{-1}$ kpc is indicated in each frame of this figure[2]. Previous studies on each of the chosen systems fully vindicates their merger classification (see appendix).

While projection effects will lead to some ambiguity in the exact ordering of systems along the sequence, the general ordering proposed by Toomre is very well supported by both observational and numerical work. To avoid the ambiguity introduced by projection effects, we break the sequence up into *early, intermediate,* and *late* stages of merging, defined optically as follows:

- Early-stage mergers are well separated and their disks are disturbed but not totally disrupted.

- Intermediate-stage mergers have distinct nuclei embedded within a common envelope of luminous material.

- Late-stage mergers have two tidal appendages emanating from a single nucleus surrounded by a mostly relaxed stellar profile.

While projection effects may cause different orderings between objects of the same stage, it is unlikely that they would lead to confusion between objects at different stages. We now describe the optical morphology of the five systems, according to their evolutionary stage.

---

[1] One should note that this "ongoing" sequence likely spans time-scales of over a Gyr.
[2] All length, mass, and luminosity scales in the following are expressed in units involving the scale factor $h$ defined in terms of the Hubble constant $H_0$ via the relationship $H_0 = 100\,h\ {\rm km\,s^{-1}\,Mpc^{-1}}$

The early-stage mergers are represented by Arp 295 (Fig. 1a) and NGC 4676 (Fig. 1b). Arp 295 consists of two disk galaxies separated by at least 95 $h^{-1}$ kpc. There is a luminous tail emerging from the southern nearly edge-on disk (Arp 295a) that reaches 60 $h^{-1}$ kpc to the south, and a stellar bridge connecting this galaxy to its northern neighbor (Arp 295b). The northern galaxy sports no such narrow tidal features, but has a plume of faint starlight extending 35 $h^{-1}$ kpc to the east. The tail and bridge indicate that significant tidal braking of the relative orbits is taking place, and it is therefore likely that the two galaxies will spiral ever closer together and eventually merge.

NGC 4676 is composed of a northern edge-on disk (NGC 4676a), and a southern inclined disk (NGC 4676b). This system is possibly at a slightly more advanced stage than Arp 295, with the two disks separated by less than one optical diameter in projection, but still distinct. The tails are of high surface brightness and both have projected lengths of 40 $h^{-1}$ kpc. Material is being exchanged between the two progenitors as is clear from the presence of a bright optical bridge between them.

The intermediate-stage is represented by NGC 520 (Fig. 1c). The two nuclei are revealed by 2$\mu$m CCD imagery (Stanford & Balcells 1990): a primary nucleus coincident with the center of the dust lane, and a secondary nucleus lying 40″ (4.4 $h^{-1}$ kpc) to the northwest. The main body is very distorted, with a bright optical tidal tail reaching 18 $h^{-1}$ kpc to the south. The northern "tail" drawn by Toomre is actually the edge of a disk (see §A-3). In deep optical imaging (Stockton & Bertola 1980; Fig. 2c), a broad plume of faint light seems to emanate from the southern tail, stretching towards the dwarf galaxy UCG 957 which lies 40 $h^{-1}$ kpc to the north of NGC 520.

The late-stages are represented by NGC 3921 (Fig. 1d) and NGC 7252 (Fig. 1e). Both systems have $r^{1/4}$ radial light profiles (Schweizer 1982, 1995, Stanford & Bushouse 1991, HGvGS) and obey the Faber-Jackson relationship (Lake & Dressler 1986) that is characteristic of normal ellipticals (Faber & Jackson 1976). The pair of tidal appendages in NGC 3921 consist of a northern plume resembling a tail bent back upon itself, reaching a radius of 30 $h^{-1}$ kpc, and a southern tail with a length of 45 $h^{-1}$ kpc. NGC 7252 was the last object presented by Toomre, and has a pair of tails stretching 48 and 70 $h^{-1}$ kpc to the east and northwest, respectively.

## 3. Observations & Data Reduction

The observed wavelengths were chosen to best address the question of the fate of gas in disk galaxy mergers. The data consist of H I spectral-line mapping using the Very Large Array[3] (VLA), and optical broad-band $R$ and narrow band H$\alpha$ images obtained at the Kitt Peak National Observatory (KPNO) and Cerro Tololo Interamerican Observatory[4] (CTIO). Tables 1 & 2 list the relevant instrumental and observational parameters in all observed bands for each system.

---

[3] The VLA of the National Radio Astronomy Observatory is operated by Associated Universities, Inc., under cooperative agreement with the National Science Foundation.

**FIGURE 1:** *R*-band CCD mosaics of the merging sequence in negative greyscale. This and all subsequent images should be viewed after rotating 90° clockwise, such that North is up and East is to the left. A horizontal black bar is drawn at a fiducial linear scale of 20 $h^{-1}$ kpc. The galaxies are arranged according to their evolutionary stage in the merging process. On the left, from bottom to top we show **(a)** the early stage merger Arp 295a (south) and Arp 295b (north). Both the bridge and the southern tail emanate from the southern galaxy. **(b)** NGC 4676a (north) and NGC 4676b (south), "The Mice", another early stage merger. **(c)** NGC 520, an intermediate stage merger. The saturated portion of this frame contains the two merging nuclei of the progenitors. A faint plume of light is seen emerging from the southeastern portion of this system towards the dwarf galaxy UCG 957 in the northwest corner. In the right column, starting at the bottom we present **(d)** NGC 3921, a late stage merger. Even though this system has a single relaxed nucleus, the southern loop leading up to the plume of light to the northeast and the southern tail are evidence of its dual-disk origin. And **(e)** NGC 7252, the "Atoms-for-Peace" galaxy. As in NGC 3921, the two extensive tidal tails point to its merger origin.

### 3.1 Radio Observations

The H I observations consist of data collected between 1989 and 1992 with the VLA in its spectral-line mode. Each galaxy was observed in two configurations: the 1.3 km D-array configuration and the 3 km C-array configuration (NGC 7252 was observed instead in the C-array configuration with the antennas of the north arm in the 10 km B-array configuration to improve the North-South angular resolution, thus compensating for its low declination). The difference in baseline lengths gives a factor of 3 difference in spatial resolution and a factor of 9 in surface brightness sensitivity between the two arrays. The field-of-view is set by the 25m diameter of the individual array elements, and amounts to 30′ full width at half maximum (FWHM). A description of the VLA is given by Napier *et al.* (1983).

The velocity channel width ($\Delta v$) was chosen to be as small as possible with enough bandwidth to cover the velocity spread of the gas seen in preexisting H I or CO single dish observations. On-line Hanning smoothing was applied to the data after which every other channel was discarded, leaving a set of independent channels, or a "velocity cube". Instrumental parameters and resulting sensitivities for the C, D, and combined C+D array data-sets are listed in Table 1. The C+D data set was constructed by adding the data in the UV plane, and will be used primarily throughout this paper. Unless otherwise noted, the "natural" weighting scheme has been used to maximize sensitivity to low surface brightness features.

Standard VLA calibration procedures were followed. The continuum was subtracted by making a linear fit to the visibility data of a range of line free channels in either side of the band. Images were made using natural weighting and the channels which showed line emission were CLEANed to remove the antenna sampling pattern (Clark 1980). The noise in the resulting images was close to thermal, *i.e.* no significant artifacts were present.

Since the data is composed of several independent observations of each system, we perform three tests to check the reality of each feature in the final images: (1) each 8 hour C-array observation is checked against the others; (2) each 4 hour D-array observation is checked against the other; (3) the C-array observations are spatially smoothed by a factor of 3 and compared with the D-array observations (failure to match features in this last test is not conclusive, since the C-array is less sensitive than D-array to scales greater than 7′ due to lack of short spacings). After careful viewing of all the data, regions containing emission were masked in the velocity cube, and a moment analysis is performed using the cutoff technique described in Bosma (1981). The output from this analysis is an integrated intensity

---

[4] CTIO and KPNO of the National Optical Astronomy Observatories are operated by the Association of Universities for Research in Astronomy, Inc., under cooperative agreement with the National Science Foundation.

TABLE 1. VLA Observing Parameters.

|                                      | A295           | N4676          | N520           | N3921          | N7252          |
|--------------------------------------|----------------|----------------|----------------|----------------|----------------|
| Phase Center:                        |                |                |                |                |                |
| $\alpha_{1950}$                      | $23^h39^m19.7^s$ | $12^h43^m42.0^s$ | $01^h21^m59.6^s$ | $11^h48^m28.9^s$ | $22^h17^m57.9^s$ |
| $\delta_{1950}$                      | $-03°54'02''$  | $+31°00'00''$  | $+03°31'53''$  | $+55°21'28''$  | $-24°55'50''$  |
| Velocity Center (km s$^{-1}$)        | 6858           | 6750           | 2260           | 5838           | 4749           |
| Observation Date:                    |                |                |                |                |                |
|   D-array                  | 8/92           | 5/91           | 5/91           | 5/91           | 12/89          |
|   C-array                  | 5/92           | 5/92           | 12/90          | 1/91           | 10/90[a]       |
| Time on Source (hrs):                |                |                |                |                |                |
|   D-arry                   | 6.0            | 6.0            | 7.0            | 7.5            | 7.0            |
|   C-array                  | 19.5           | 21.5           | 21.5           | 21.5           | 22.5           |
| Bandwidth (MHz):                     |                |                |                |                |                |
|   D-arry                   | 6.25           | 6.25           | 3.13           | 3.13           | 6.25           |
|   C-array                  | 6.25           | 6.25           | 3.13           | 3.13           | 3.13           |
| Number of Channels:                  |                |                |                |                |                |
|   D-arry                   | 63             | 32             | 63             | 63             | 32             |
|   C-array                  | 63             | 63             | 63             | 63             | 63             |
| Channel Separation (km s$^{-1}$):    |                |                |                |                |                |
|   D-arry                   | 21.6           | 43.1           | 10.5           | 10.7           | 42.5           |
|   C-array                  | 21.6           | 21.5           | 10.5           | 10.7           | 10.6           |
| Map Characteristics (Natural Weighting): |            |                |                |                |                |
| Integrated C+D Flux                  |                |                |                |                |                |
|   (Jy km s$^{-1}$)         | 12.0           | 3.77           | 32.9           | 3.94           | 3.75           |
| Synthesized Beam FWHM:               |                |                |                |                |                |
| (Major Axis × Minor Axis)            |                |                |                |                |                |
|   D-arry                   | $68''\times54''$ | $64''\times61''$ | $66''\times54''$ | $60''\times59''$ | $96''\times52''$ |
|   C-array                  | $21''\times19''$ | $19''\times17''$ | $19''\times18''$ | $19''\times17''$ | $21''\times16''$ |
|   C+D                      | $27''\times25''$ | $20''\times19''$ | $24''\times24''$ | $24''\times21''$ | $27''\times16''$ |
| Noise Level (1$\sigma$):             |                |                |                |                |                |
| — rms noise (mJy beam$^{-1}$)        |                |                |                |                |                |
|   D-arry                   | 0.38           | 0.31           | 0.53           | 0.50           | 0.37           |
|   C-array                  | 0.30           | 0.25           | 0.31           | 0.28           | 0.32           |
|   C+D                      | 0.22           | 0.16           | 0.27           | 0.22           | 0.20           |
| — rms noise ($10^{19}$ cm$^{-2}$)    |                |                |                |                |                |
|   D-arry                   | 0.3            | 0.4            | 0.2            | 0.2            | 0.3            |
|   C-array                  | 1.8            | 1.9            | 1.0            | 1.0            | 1.1            |
|   C+D                      | 0.8            | 2.0            | 0.5            | 0.5            | 2.2            |

Notes to TABLE 1

[a] NGC 7252 used the hybrid CnB-array instead of C-array. This array has a long north arm, which is particularly useful for southern sources such as NGC 7252.

map (zeroth moment), an intensity-weighted velocity map (first moment), and an intensity-weighted velocity dispersion map (second moment). The cleaned channel maps, the zeroth and first moment maps, and a detailed description of the data for each system are presented in the appendix. Various supplementary maps are given in Hibbard (1995), and are available upon request.

TABLE 2. CTIO/KPNO Observing Parameters.

| System: | A295, N520 | N4676, N3921 | N7252 | |
|---|---|---|---|---|
| Date | Oct 92 | Mar 92 | Sep 90 | Aug 91 |
| Telescope | KPNO 0.9m | KPNO 2.1m | CTIO 4m | KPNO 2.1m |
| Detector | T2kA | T1kA | TI 800 | ST1k |
| Readout Mode | binned 1×1 | binned 2×2 | binned 2×2 | binned 2×2 |
| Focal Ratio | f/7.5 | f/7.5 | f/2.7 | f/7.5 |
| Field of View: | | | | |
| — Single CCD frame | $23.'4 \times 23.'4$ | $5.'2 \times 5.'2$ | $4.'0 \times 4.'0$ | $4.'6 \times 4.'6$ |
| — Final Image | $35' \times 29'$, $25' \times 25'$ | $10' \times 10'$, $11' \times 11'$, | $12' \times 8'$ | $12' \times 8'$ |
| Pixel Size | $0.''687$ | $0.''608$ | $0.''600$ | $0.''533$ |
| Seeing[1] | $2.''4$, $2.''8$ | $1.''9$, $1.''9$ | $1.''2$ | $2.''0$ |
| Sky Brightness (mag arcsec$^{-2}$) | 20.0, 20.1 | 20.4, 20.4 | 20.5 | — |
| Sky Noise[2] (mag arcsec$^{-2}$) | | | | |
| — $1\sigma$ (1×1) | 26.0, 25.9 | 27.0, 26.4 | 26.9 | — |
| — $3\sigma$ (9×9) | 27.2, 27.1 | 28.2, 27.6 | 28.1 | — |
| Filters | $R$, H$\alpha$ | $R$, H$\alpha$ | $R$ | $R$, H$\alpha$ |
| $\lambda_0$(H$\alpha$) | 6689 Å, 6619 Å | 6693 Å | — | 6649 Å |
| $\Delta\lambda$(H$\alpha$) | 78 Å, 67 Å | 81 Å | — | 70 Å |
| Effective Exposure Time[3] | | | | |
| — $R$ | 7×600s, 5×600s | 20×300s, 15×300s | 4×100s | 300s |
| — H$\alpha$ | 4×900s | 3×900s | — | 3×1200s |

Notes to TABLE 2

[1] Typical size of star, FWHM, on final mosaic image.

[2] $R$-band Noise calculated in original frame (1x1) or after 9×9 boxcar median filter is applied to fainter levels. See text.

[3] The approximate number of frames that contribute to the optical features times the exposure time of an individual frame. More frames were actually taken to cover extended H I features or nearby galaxies, but not all of these contain the optical features of interest.

### 3.2 Optical Observations

The optical data were obtained with a CCD camera in four observing runs at either CTIO or KPNO between 1990 and 1992. Table 2 lists the details of the observations and the resulting sensitivities.

A step-and-stare technique was used, in which the telescope is moved a fraction of an arcminute between individual, short exposures. Generous overlap between the different exposures ensures that each portion of the galaxy is imaged many times. Extreme care is taken to flat-field the exposures, using dome-flats and dark sky-flats, and to monitor and correct for temporal variations in the night sky brightness. A mosaic image is then constructed from all the images in a given filter using an extension of the technique developed by Guhathakurta & Tyson (1989). The individual frames are registered to one another to $\lesssim 0.''3$ by using the common centroids of approximately 20 unsaturated stars. Finally, absolute astrometry of each optical image was derived using four or more stars in the Hubble Space Telescope Guide Star Catalog.

The data were photometrically calibrated during the fall 1990 and 1992 observing runs. Stars chosen from the Landolt UBVRI extension (Landolt 1983) were observed in each filter at a range of airmasses during photometric conditions. Each system was observed on the same night except for NGC 4676, which is not up in the fall. This system had to be calibrated off of the observations of NGC 3921 taken during the same observing run (spring 1992), which in turn was calibrated directly in the fall 1992 run. The zero-points and color terms reproduce the colors of the standard stars to within 0.015 mag. The observed colors of the galaxies were corrected for Galactic extinction using the $A_B$ values of Burstein & Heiles (1984) and the interstellar reddening curve from Johnson (1965). K-corrections were made according to the prescription in the Second Reference catalog of Bright Galaxies (hereafter RC2, de Vaucouleurs *et al.* 1976). Due to the "peculiar" classification of these galaxies, no corrections were made for internal absorption.

Comparison of our photometry with published data on Arp 295 (Beaver *et al.* 1974, Schombert *et al.* 1990) and NGC 520 (Krienke 1975, Stanford & Balcells 1990) gives agreement to better than 0.1 mag. We adopt this as the overall uncertainty of the absolute photometry. The zero point for NGC 4676, which was not calibrated directly, remains more uncertain — perhaps by as much as 0.3 mag. The accuracy of the flat-fielding and the high degree of matching of the individual frame edges in the final mosaic image assure us that the sky background varies by less than 0.1% across the mosaic.

A narrow-band H$\alpha$ filter centered near the red-shifted H$\alpha$ line for each system was used. The filter pass-band is about 70 Å wide and thus includes emission from the [N II] $\lambda\lambda$ 6548 Å, 6584 Å lines (rest wavelengths). In order to subtract the continuum, broad-band $R$ exposures of 300 s each were obtained immediately before or after the H$\alpha$ frames. The H$\alpha$ data for Arp 295 and NGC 520 were obtained under photometric sky conditions and were calibrated via observations of spectrophotometric standards (Massey *et al.* 1988) taken on the same night, and converted to flux units using the Hayes and Latham (1975) calibration (see Massey *et al.* 1988). The calibration for NGC 4676 was kindly provided by J.C. Mihos (from Mihos *et al.* 1993), and that for NGC 3921 was taken from the aperture photometry of Kennicutt (1992a). The H$\alpha$ data for NGC 7252 remain uncalibrated.

In order to arrive at an image with very low isophotal detection limits without loss of detail on the brighter regions, combined multi-resolution images were constructed. Features with an $R$-band surface brightness $\mu_R \gtrsim 23$ mag arcsec$^{-2}$ were left at full resolution, features with surface brightness from 23 to 25 mag arcsec$^{-2}$ were smoothed with a 3-pixel boxcar median filter, and pixel values fainter than this were smoothed with a 9-pixel boxcar. This results in sensitive images of the faint optical light reaching a surface brightness limit of $\mu_R \gtrsim 27$ mag arcsec$^{-2}$ ($3\sigma$; see Table 2).

To measure the luminosity and extent of the faint starlight, foreground Galactic stars and background galaxies are removed from the broad-band images. Point spread function subtraction techniques could not be utilized, since the construction of the mosaic image results in a PSF which varies non-monotonically with location on the final image. Instead, regions containing and surrounding stars are replaced with a local determination of the background, and the resulting image was smoothed as above. This image will be referred to in the following as the deep star-subtracted image. The replacement process results in some systematic errors in measuring the extent of the faint isophotes. To estimate these errors, we take all pixels which have been replaced with a locally determined background and assign them a 25% error, except at regions where the galactic light profile changes rapidly. The latter are assigned a 50% error. These values are not reduced by smoothing. The errors on all other pixels are assigned according to poisson statistics, detector noise, and the appropriate smoothing factor. These errors are used to estimate the uncertainties in the luminosity measurements, which we find to be good to within ∼10%. We adopt this as a standard error on all reported luminosities in this paper. For NGC 4676, there are no galaxies or very bright stars intruding within the faint light, and the observations seem believable down to their statistical limit, which is the deepest presented here ($\mu_R(3\sigma) = 28.2$ mag arcsec$^{-2}$).

## 4. Results

In this paper we wish to stress the observational trends along the proposed merger sequence. For this purpose, we present the main results of this work as a series of montages, given in Figures 2–6, with the galaxies ordered according to their evolutionary stage in the merging process. The appendix presents detailed descriptions and supplementary figures individually for each system. In this and the next section we describe these data much more generally.

Figure 2(a–e) presents a high contrast version of the deep, multi-resolution $R$-band image of each system in negative greyscale. The faintest light levels in this figure are of order $\mu_R \gtrsim 27$–28 mag arcsec$^{-2}$. Figure 3(a–e) [PLATE 1] presents both cold and warm gas phases — with the neutral hydrogen distribution in blue and the ionized hydrogen in red — upon the deep $R$-band mosaic image of the starlight in green and white. Figure 4(a–e) shows the deep star-subtracted image described in §3.2 contoured at $\mu_R = 23.5$, 25.0, and 26.5 mag arcsec$^{-2}$ drawn upon a greyscale of the H I distribution. The H I column densities range from $2\times10^{19}$ cm$^{-2}$ (white) to $1\times10^{21}$ cm$^{-2}$ (black). Figure 5(a–e) [PLATE 2] presents the intensity-weighted H I velocity field for each system, with the sense of velocity indicated by the appropriate color (red-shifted or blue-shifted with respect to the systemic velocity). All velocities in this paper are heliocentric. In figure 6(a–f) we present a close-up of the central regions of each system, showing details of the ionized gas distribution. The $R$-band image is reproduced in negative greyscales, while the contours upon a white background indicate the continuum subtracted H$\alpha$+[N II] emission.

The global properties of these systems are tabulated in Table 3. The first line gives the reference listed in the table notes for the optical, CO, and IR information. Line (2) lists the Hubble classification, taken from the RC2; line (3) lists the adopted distance to the system; line (4) lists the corrected optical diameter at $\mu_B = 25$ mag arcsec$^{-2}$ ($D_{25}$) from the RC2; line (5) lists the adopted inclination from the reference in the notes. These systems are all sufficiently disturbed that the ratio of the major to minor axis diameters is not a good measure of their inclination. For three systems the inclination is based on a dynamical model (Arp 295, NGC 4676, NGC 520). For NGC 7252, the inclination is based on the shape of the inner H$\alpha$ emission. For NGC 3921, the ionized gas exhibits chaotic rather than rotational motion (Schweizer 1995), and so no inclination is given. Line (6) of Table 3 lists the far-infrared luminosity in the 42.5–122.5 $\mu$m band ($L_{\rm FIR}$) derived from the 60$\mu$m and 100$\mu$m IRAS measurements (Helou *et al.* 1988). Since NGC 4676 is not resolved by the IRAS beam, we take advantage of the tight correlation between IR and radio continuum flux (Condon *et al.* 1991) and split the total IR luminosity between the component systems in proportion to their contribution to the total radio continuum flux (line 17).

Lines 7–11 list the optical properties of each system, taken from the listed references and the present work. In line (7) we give the apparent corrected blue magnitude ($B_T^o$) of each system with the corresponding blue luminosity ($L_B$) given in line (8). Line (9) gives the measured $R$-band luminosities ($L_R$), corrected for Galactic absorption and red-shift as described in §3.2. Lines (10) and (11) list the optical heliocentric velocity ($V_{opt}$) and velocity linewidth ($\Delta W_{opt}$) derived from emission line measurements. Line (12) lists the central H$\alpha$ luminosities taken from the literature (usually measured with an aperture of order 5″), while line (13) gives the total measured H$\alpha$+[N II] luminosities from the present observations.

Lines 14–16 list the single dish $^{12}$CO(1→0) information from the reference listed in the table notes: line (14) lists the central velocity of the $^{12}$CO(1→0) line emission ($V_{CO}$), line (15) gives the $^{12}$CO(1→0) velocity width ($\Delta W_{CO}$), and line (16) lists the derived H$_2$ gas mass[5] ($M_{\rm H_2}$).

Lines 17–20 list the results from the present H I observations. Line (17) list the 1.4 GHz radio continuum flux ($S_{\rm 1.4GHz}$) in mJy. Line (18) gives the best estimate of the systemic velocity from the H I kinematics, and line (19) lists the H I velocity width ($\Delta W_{HI}$). Line (20) lists the total H I gas mass[6] ($M_{\rm HI}$).

---

[5]  Molecular hydrogen column densities $N_{\rm H_2}$ and masses are calculated from the integrated $^{12}$CO(1→0) emissivity $\int T_b dv$ (where $T_b$ is the brightness temperature in K and the integral is over the line, with the velocity $v$ expressed in km s$^{-1}$) using a conversion factor $\alpha_{\rm CO}$ such that: $N_{\rm H_2} = 4 \times 10^{20} \alpha_{\rm CO} \int T_b dv$ cm$^{-2}$. Best estimates of $\alpha_{\rm CO}$ range from 0.5 to 1.5 under most conditions (Maloney 1989).

[6]  Neutral hydrogen masses are calculated from the integrated H I flux $\int S_{\rm HI} dv$ (where $S_{\rm HI}$ is the continuum subtracted H I line specific intensity in Janskys and the integral is over the line, with the velocity $v$ expressed in km s$^{-1}$), assuming that the atomic gas is optically thin, according to the equation $M_{\rm HI} = (2.36 \times 10^5 M_\odot) \times D_{\rm Mpc}^2 \times \int S_{\rm HI} dv$, where $D_{\rm Mpc}$ is the distance to the source in $h^{-1}$ Mpc.

**FIGURE 2:** A high contrast version of the $R$-band mosaic images presented in Fig. 1 shown to emphasize the faint outer features. The faintest light levels in this figure are $\mu_R \gtrsim 27\text{–}28$ mag arcsec$^{-2}$. Left column, bottom to top: **(a)** Arp 295a&b; **(b)** NGC 4676a&b; **(c)** NGC 520 & UCG 957. Right column, bottom to top: **(d)** NGC 3921 and **(e)** NGC 7252. Note that the bright features in Fig. 1 have extensive halos of faint light, but that no new features are revealed.

**FIGURE 3:** A false-color composite image of the merger sequence, with the neutral hydrogen distribution shown in blue, the continuum-subtracted ionized hydrogen distribution shown in red, and the starlight from the $R$-band mosaic shown in yellowish-green and white. Bright stars that were saturated on the CCD appear red in this image. In the top row, from left to right: **(a)** Arp 295a&b; **(b)** NGC 4676a&b; and **(c)** NGC 520 and UCG 957. Bottom row, from the left, **(d)** NGC 3921; and **(e)** NGC 7252.

**FIGURE 4:** The deep star-subtracted image contoured at $\mu_R = 23.5$, $25.0$, & $26.5$ mag arcsec$^{-2}$ drawn upon a greyscale of the H I column densities, ranging from $2\times10^{19}$ cm$^{-2}$ (white) to $1\times10^{21}$ cm$^{-2}$ (black). Left: **(a)** Arp 295a&b. Note the diffuse halo of light around Arp 295a, the gap in the gas distribution along the bridge, and the optical extension to the S tail. **(b)** NGC 4676a&b, with a diffuse halo and an optical extension to the tails. **(c)** NGC 520 & UCG 957, showing a striking anticorrelation between the outer gas and light. Right: **(d)** NGC 3921, with its gas-rich southern tail and gas-poor northern plume. And **(e)** NGC 7252, with a very low central atomic gas content. Note the decreasing prominence of the darkest H I greyscale along the sequence and the difference between the lowest light contour and the lightest H I greyscale.

**FIGURE 5:** The intensity-weighted H I velocity field for each system, with the sense of velocity (red-shifted or blue-shifted with respect to the systemic velocity) indicated by the appropriate color. The systemic velocities appear either as green or yellow. In the following we list the heliocentric systemic velocities and velocity extrema, in km s$^{-1}$, after the system identification in the format ($V_{min}, V_{sys}, V_{max}$): Along the top row, from left to right: **(a)** Arp 295, (6500, 6880, 7150); **(b)** NGC 4676, (6300, 6600, 6900); **(c)** NGC 520, (2050, 2260, 2450). Along the bottom row are **(d)** NGC 3921, (5780, 5895, 5970); and **(e)** NGC 7252, (4600, 4740, 4880). The smooth runs of velocity support a tidal origin for the outer features.

**FIGURE 6:** Close-up of the central regions of each system, showing details of the ionized gas distribution. The $R$-band image is reproduced in negative greyscales, while the contours upon a white background indicate the continuum subtracted H$\alpha$+[N II] emission. Along left hand side we have **(a)** Arp 295b, **(b)** Arp 295a; **(c)** NGC 4676a&b; On the right we have **(d)** NGC 520; **(e)** NGC 3921; and **(f)** NGC 7252. Note the minor axis plumes seen in **(a)**, **(c)**, and **(d)**, and the offset bar in **(c)**.

TABLE 3. Global Properties[1] of Systems.

Notes to TABLE 3

[1] All length, mass, and luminosity scales in the following are expressed in units involving the scale factor $h$ defined in terms of the Hubble constant $H_0$ via the relationship $H_0 = 100\,h\,\mathrm{km\,s^{-1}\,Mpc^{-1}}$

[2] The infrared luminosity in the 42.5–122.5 $\mu$m band, calculated from the 60$\mu$m and 100$\mu$m IRAS fluxes using the prescription given in Helou *et al.* 1988. This luminosity is expressed in terms of the bolometric luminosity of the Sun, $L_\odot = 3.83 \times 10^{33}$ ergs s$^{-1}$.

[3] The $B$-band fluxes are converted to solar values by adopting an absolute magnitude in the $B$-band of the sun of 5.48 (Mihalas & Binney 1981). As such, the unit of blue luminosity is actually $L_{\odot,B}$, the luminosity of the Sun in the $B$-band, rather than the bolometric luminosity of the Sun. We retain these units for direct comparison with previous work, and will use the symbol L$_\odot$ for simplicity.

[4] The $R$-band fluxes are converted to solar values by adopting an absolute magnitude in the $R$-band of the sun of 4.31 (Mihalas & Binney 1981 and Johnson 1966). See note [3].

Notes to TABLE 3 (cont'd)

[5] These rather confusing units arise because the IR luminosity is expressed in terms of the Sun's bolometric luminosity, while the $B$-band luminosity is expressed in terms of the Sun's luminosity in the $B$-band.

[a] A295 $B_T^o$ and $L_{H\alpha}$ from Keel *et al.* 1985; $V_{opt}$ and $\Delta W_{opt}$ from Stockton 1974b; inclination from Toomre & Toomre 1972; $^{12}$CO(1→0) data from Combes & Casoli in preparation; IR data from Surace (personal communication).

[b] N4676 $B_T^o$ from Doi *et al.* 1995; $L_{H\alpha}$ from Keel *et al.* 1985; $V_{opt}$ and $\Delta W_{opt}$ from Stockton 1974a; inclination from Toomre & Toomre 1972; $^{12}$CO(1→0) and IR data from Casoli *et al.* 1991.

[c] N520 $B_T^o$ from RC2; inclination, $V_{opt}$, and $\Delta W_{opt}$ from Bernlöhr 1993; $L_{H\alpha}$ from Young *et al.* 1988; $^{12}$CO(1→0) and IR data from Young *et al.* 1989 and Sanders *et al.* 1988a. Parenthetical H I values represent absorption-corrected quantities; see text.

[d] N3921 $B_T^o$ from Lake & Dressler 1986; $V_{opt}$ from Huchra *et al.* 1992; $L_{H\alpha}$ and $\Delta W_{opt}$ from Schweizer 1995; $^{12}$CO(1→0) data from Combes & Casoli in preparation; IR data from Deutsch & Willner 1987.

[e] N7252 $B_T^o$ from Lake & Dressler 1986; no calibrated H$\alpha$ measurement is available; $V_{opt}$ and $\Delta W_{opt}$ and inclination from Schweizer 1982; $^{12}$CO(1→0) and IR data from Dupraz *et al.* 1990.

The measurements presented in this section have been made for the main systems and their associated tidal appendages, either optical or gaseous. So, for instance, while the three small galaxies to the west of Arp 295a are also seen in H I (Fig. 3a), they appear as quite separate entities within the velocity field rather than as part of the tidal features (Fig. 5a) and are not included in the reported H I mass. For similar reasons, the galaxies to the north of Arp 295b and to the west of NGC 3921 are not included in the following, while UCG 957 is included for the measurement of the NGC 520 system, since it falls smoothly and continuously within the kinematics of the outer H I ring (Fig. 5c).

The remaining lines list ratios of the above quantities. Line (21) lists the total gas mass ($M_{gas}=M_{\rm HI}+M_{\rm H_2}$; no correction has been made for He). Line (22) lists the IR-to-blue light ratio ($L_{\rm FIR}/L_B$), believed to be a measure of the massive star formation rate (Solomon & Sage 1988). Line (23) lists the ratio of the IR luminosity to molecular gas mass ($L_{\rm FIR}/M_{\rm H_2}$), also referred to as the "star formation efficiency" as it is a measure of the number of massive stars being formed per molecular cloud (Young *et al.* 1986, Sanders *et al.* 1986, Solomon & Sage 1988). Line (24) lists the H$_2$ mass per unit blue luminosity ($M_{\rm H_2}/L_B$), line (25) lists the H I mass per unit blue luminosity ($M_{\rm HI}/L_B$), and line (26) lists the molecular to atomic gas mass ratio ($M_{\rm H_2}/M_{\rm HI}$).

In the following we quote percentages of gas and light separately for the "inner" and "outer" regions of each system. These regions cannot be delineated by a constant light isophote, since outer features (*e.g.* tails, loops) are often projected across the inner regions, and the main bodies have strongly varying amounts of star formation and dust. For example both of the galaxies of "The Mice" have a standard isophote, $D_{25}$, that is twice the size of the galactic bodies seen in Fig. 1b because the base of the tails are brighter than $\mu_B=25$ mag arcsec$^{-2}$.

This Page Left for Table 4

This Page Left for Table 5

**FIGURE 7a:** The logarithm of the H I mass to $R$-band light ratios ($M_{\rm HI}/Lr$) as a function of distance measured along the tidal features for the early stage mergers Arp 295 and NGC 4676. The bars along the left hand side indicate the range of $M_{\rm HI}/L_R$ found from the sample of Wevers *et al.* (1986) measured at 0.5, 1.0, and 1.5 times the standard optical radius from the RC2. At these early stages, the ratios are typical of the outer regions of disk galaxies, and the light extends beyond the gas, leading to a sharp decrease in $M_{\rm HI}/L_R$.

Since our $R$-band photometry is of nearly equal quality for each system, we use our own judgment, aided by color and kinematic information, to define separately the inner and tidal regions (for an example of how this information is used, see HGvGS). We have attempted to chose these regions so that they would roughly correspond to an isophote of $\mu_R$=22 mag arcsec$^{-2}$ in the absence of tidal features. The inner regions so defined include the main disks and bulges of each system and are increased in size by the half-width of the H I beam. The tidal regions include only the main H I tidal features, and not the optical plumes.

The results of this decomposition are tabulated in Table 4 for the inner regions and Table 5 for the tidal features. In Table 4, line (1) lists the diameter of the inner region so delineated ($d_{opt}$), and line (2) gives the total $R$-band luminosity within this diameter. Lines 3–5 list the H I mass, total gas mass, and dynamical mass within $d_{opt}$. The dynamical mass within $d_{opt}$ has been calculated for all galaxies except NGC 3921 using the equation $M_{dyn}(d_{opt}) = 2.45 \times 10^4 \ d_{opt} \ [\frac{\Delta W}{sin(i)}]^2 M_\odot$ (Shostak

**FIGURE 7b:** As in Fig. 7a, but for the intermediate and late stage systems NGC 520, NGC 3921, and NGC 7252. In these later stages, the tidal features are progressively gas rich to the outer most regions, with $M_{\rm HI}/L_R$ ratios much higher than in the outer regions of disk galaxies.

1978), where $\Delta W$ is the maximum observed velocity width from Table 3. For NGC 3921, there is little rotational signature in the ionized gas kinematics (Schweizer 1995), and we instead calculate the dynamical mass using the equation $M_{dyn}(d_{opt}) = 2.33 \times 10^5 \ d_{opt} \ \sigma^2$ (Devereux *et al.* 1987), where $\sigma$ is the velocity dispersion.

Lines 6–8 list ratios of various quantities for the inner regions. Line (6) lists the ratio of the $H_2$ to H I mass. Line (7) lists the total gas mass to $R$-band light ratio ($M_{gas}/L_R$), and line (8) gives the fraction of the dynamical mass which can be accounted for by gas ($M_{gas}/M_{dyn}$). We find that the systems with the largest value of this last quantity exhibit minor axis plumes of H$\alpha$ emission (*cf.* Fig. 6). To emphasize this point, we indicate galaxies with such plumes by an asterisk ($^*$) above the entry on line (8).

In Table 5, line (1) lists the projected linear extent of the tidal feature ($l_{opt}$) measured at the 26.5 mag arcsec$^{-2}$ isophote of the $R$-band image. Line (2) lists the extent of the H I distribution ($l_{HI}$) measured at $5\times10^{19}$ cm$^{-2}$. The projected lengths are measured from the inner region continuously along the tidal feature to its end. The next 6 entries list the H$\alpha$+[N II] and $R$-band luminosity and H I mass of each feature, in terms of absolute luminosities and masses (lines (3), (4) and (5)) and as a percentage of the parent systems total values (lines (6), (7), and (8)). Finally, line (9) lists the H I mass to $R$-band light ratio ($M_{\rm HI}/L_R$).

As an aid in the following discussion, we present in figure 7 a plot of gas-to-light surface brightness ratios ($M_{\rm HI}/L_R$, in $M_\odot L_\odot^{-1}$) as a function of distance along the tidal features of each system. These plots were made by convolving the star-subtracted optical images by the appropriate H I beam and measuring along the spine of the tidal features. In Fig. 7a we plot the profiles for the first two systems, and in Fig. 7b we plot the other three systems. For comparison we use the field spiral sample of Wevers *et al.* (1986) which have both optical and H I surface brightness radial profiles measured (16 galaxies). For this comparison sample we plot the *full* range (and *not* the standard deviation) of $M_{\rm HI}/L_R$ at three fiducial radii: at 0.5, 1.0, and 1.5 times the standard optical radius from the RC2. These are indicated by the heavy vertical lines to the left in each plot.

## 5. Observed Trends Along the Sequence

### 5.1 A Sequence of What?

The studies cited in the appendix provide strong evidence that each of the chosen systems are mergers of disk galaxies, with Arp 295 and NGC 4676 at a relatively early stage of merging, NGC 520 in the midst of the coalescence of the progenitor nuclei, and NGC 3921 and NGC 7252 at the final stages of coalescence. The H I velocity fields (Fig. 5, see also appendix) show that projections are not confusing our classification. Therefore, it seems well established that the five systems *do* represent a sequence of merging disk galaxies with less and less time remaining until the merger is complete.

However, the sequence is not completely homogeneous. In particular, it appears that the mass ratios and Hubble types of the progenitors and the encounter geometries differ in important ways. For example, Arp 295b is experiencing a retrograde encounter, and we believe that both NGC 520 and NGC 3921 involved one gas-rich and one gas-poor progenitor (see appendix and §6.2.3). Therefore, one cannot think of the present sequence as snapshots in the time evolution of a single merging encounter.

Additionally, it is important to realize that since the selection of these objects was predicated on their having several long, bright tails, certain encounter geometries are more likely to be represented than others. In particular, this criterion favors mergers resulting from a direct or prograde encounter (*i.e.*, with the progenitors disks rotating in the same sense as their mutual orbit) of intermediate impact parameter. Long and robust tidal features can only be created if the force acts for a long enough period of time (favoring prograde encounters, TT; White 1979) and is strong enough (favoring small impact parameters, Olson & Kwan 1990, Barnes 1992). Although the impact parameter must be small enough to produce a strong response in the disk, it should not be too small, as the galaxies would then merge before long tails had time to develop (Barnes 1992). Further, the presence of significant tails requires the progenitors to have fairly modest dark matter contents ($M_{dark}/M_{luminous} < 10$, Dubinski *et al.* 1995).

This is the framework from which we interpret the H I and H$\alpha$ data: that the sequence represents progressively advanced mergers between disk galaxies of comparable mass ($M_1/M_2 > 0.3$). It is possible that only certain encounter geometries are represented, though only explicit modeling of the data will establish this. Given these caveats, we now attempt to identify the phenomena which appear to be associated with either the early, late, or all stages of merging, in the hopes of identifying the important processes at play.

### 5.2 Optical Morphology

Since the members of this evolutionary sequence were chosen by their optical morphology, the observed trends in the starlight are those described in §2. In the early stages the light is primarily within separate disk galaxies, while in the later stages the light emanates primarily from a single relaxed remnant with an $r^{1/4}$ light profile. Fig. 2 shows that the faint starlight is widespread at all stages, and that no new optical features are uncovered by the deep imaging: the fainter light appears as extensions to the brighter features already visible in Fig. 1. The often patchy looking tidal features seen in shorter exposure images fill in with light on the deep images.

### 5.3 Plumes of Ionized Hydrogen

In the earlier stages of merging there are often remarkable plumes and arcs of ionized gas emanating from the nuclear regions along the minor axis, see *e.g.* Arp 295a (Fig. 6a), NGC 4676a (Fig. 6c) and NGC 520 (Fig. 6d). The systems which show such plumes also have the highest ratio of gas mass to dynamical mass in the central region ($M_{gas}/M_{dyn}$; line 8 of Table 1): Arp 295b has an inner gas fraction that is five times that of its southern neighbor (0.11 $h^{-1}$ *vs.* 0.02 $h^{-1}$); NGC 4676a has nearly three times the central gas fraction of NGC 4676b (0.08 $h^{-1}$ *vs.* 0.03 $h^{-1}$); and NGC 520 has $M_{gas}/M_{dyn} > 0.14\ h^{-1}$. No other quantity shows such a differentiation between systems with and without minor-axis plumes.

In the early stages, star formation is spread throughout the disks and has an irregular morphology with many knots. In the later stage mergers, the inner H II emission is concentrated in the inner few arcseconds and appears more regular, without any significant substructure. The irregular inner H$\alpha$ features of NGC 3921 and NGC 7252 are likely only in projection close to the center, as their association with the neutral hydrogen in these areas suggests that they are confined to the outer tidal tails and loops (*cf.* HGvGS for NGC 7252).

### 5.4 Central Atomic Gas Content

The most obvious difference in the gas distribution along the sequence is the lack of high column density H I in the later stages, as evidenced by the decrease in prominence of the darkest greyscales from Fig. 4a through Fig. 4e. This figure further shows that the high column density gas in the earlier stages is associated with large galactic disks, while in the later stages these disks are conspicuously lacking (except for tiny molecular-gas disks in the center), and the gas is relegated to the tidal features.

**FIGURE 8:** Fraction of light and atomic gas contained within the inner regions along the sequence. The early stage mergers are at the far left of this plot, and the late stage mergers are at the far right. We see that while the fraction of light contained within the remnant body is relatively constant, the fraction of atomic gas in these regions drops precipitously for the late stage mergers. The range of $M_{\rm HI}$ values shown for NGC 520 indicate the correction for central H I which is missed due to the central absorption.

As a result, we find a decrease in the central atomic gas content along the sequence. This is quantified in figure 8, where we have plotted the fraction of light and atomic gas contained within the inner regions. Recall that these regions have been selected to approximate the $\mu_R \sim 22$ mag arcsec$^{-2}$ optical isophote in the absence of tidal features (§4). This plot shows that while the fraction of light contained within the remnant body is relatively constant, the fraction of the atomic gas that is found within these regions is much higher in the systems which still have significant disks (Arp 295, NGC 4676, NGC 520) than in the systems with $r^{1/4}$ light profiles (NGC 3921, NGC 7252). We therefore conclude that not only do some mergers rearrange the light profiles of the progenitor disk galaxies into $r^{1/4}$ profiles, but in so doing also rid the progenitors of much of their atomic gas supply. The only atomic gas which survives is that which has been ejected into the dynamically "cold" tidal tails.

### 5.5 H I vs. Optical Tidal Morphologies

Examining Fig. 4, one finds progressively less correspondence between the lowest optical contour and the faintest gas greyscale along the sequence. In the early stages, the differences are subtle: the edges of the optical tails extend slightly further than the gas; there are some misalignment. But at the later stages the differences are stark: there are extensive optical features without gaseous counterparts, and the gaseous tails extend far beyond the end of their optical counterparts.

We find it convenient to separate the extra-nuclear light into three different categories: *tails, plumes*, and *halos*. Following Schombert *et al.* (1990), we define tails as having gaussian shaped intensity cross-sections and plumes as plateau shaped[7]. Since the plumes in both this work and the Schombert *et al.* (1990) survey are not in general fainter than the tails, we conclude that they all cannot be simply tails viewed face-on. By halos we refer to the faint, featureless extensions of light found around brighter features, such as bulges, disks, tails, and plumes (which are not featureless). While this distinction may seem arbitrary or unphysical, we find that it also separates the lower surface brightness features according to their H I properties:

- Optical tails are gas rich (Arp 295a, NGC 4676a&b, NGC 520, NGC 3921, NGC 7252). However, there are often displacements between the optical and gaseous peaks (Arp 295a, Arp 295 bridge, NGC 520 S tail) and gaps in the gas distribution (Arp 295 bridge, NGC 4676b, N7252). When displacements are present, they are in the sense that the gas lies behind the optical light in terms of its orbital motion.

- Optical plumes tend to be gas poor (NGC 520, NGC 3921).

- Optical halos are gas poor (Arp 295a, NGC 4676, NGC 3921, NGC 7252).

- In the early stages, the optical tails extend a little beyond the ends of the gaseous tails (Arp 295a, NGC 4676a&b), while at the later stages, the gaseous tails are much longer than the optical tails (NGC 520, NGC 3921, NGC 7252).

- The peaks in the outer gaseous extensions are anti-correlated with any nearby optical features (NGC 520 plume, NGC 3921 S loop).

---

[7] Note that this definition of plume differs from that used to describe the minor-axis H$\alpha$ emission, which has nothing to do with intensity profiles.

## 5.6 Extent of Tidal Features

One of the more striking impressions from these figures is that large quantities of gas and stars are seen at large radii. Table 2 shows that on average $\sim 2\times10^9\,h^{-2}\,M_\odot$ of atomic gas and $\sim 3\times10^9\,h^{-2}\,L_\odot$ of starlight for each system are sent to these distant regions. We find that the tidal H I reaches from 2 (NGC 4676a) to 7.5 (NGC 7252) times the systems standard blue radius (one-half of $D_{25}$). For comparison, a sample of field spirals taken from Table 4 of Cayatte *et al.* (1994; compiled from the work of Warmels 1986, Wevers 1986, and Broeils 1994) have radii that reach from 0.6 to a maximum of 4.6 times $\frac{1}{2}D_{25}$ at the last measured H I point, with a mean of 2.1±0.7.

To further investigate the physical extent of these systems, we plot the intensity profiles for both stars and gas in Figure 9a&b. The intensity level is plotted along the $x$-axis ($\mu_R$ for the optical surface brightness and $N_{HI}$ for the H I column density), and the area subtended by material at or above a given intensity level is plotted along the $y$-axis.

We use several sources to gauge the optical extent of "normal" disk galaxies: (1) a sample of face-on undisturbed disk galaxies from de Jong & van der Kruit (1994), plotting the radius corresponding to $\mu_R$=23.5 mag arcsec$^{-2}$ and at the faintest measured isophote (open and filled squares in Fig. 9a); (2) a sample of edge-on disk galaxies compiled from the studies of van der Kruit and Searle (1982; open hexagons in Fig. 9a) and Barteldrees & Dettmar (1994; filled hexagons in Fig. 9a), measured at the "edge" of the optical light; and (3) the sample of Wevers *et al.* (1986), plotting the radius corresponding to $\mu_B$ =25 mag arcsec$^{-2}$ (plotted against the corresponding $\mu_R$ value calculated from the published $\mu_J$ and $\mu_F$ values using the transformation equations given in Wevers *et al.* 1986), and at the faintest measured isophote (open and filled triangles). These three samples give us 132 comparison points at the brighter isophote, and 100 points at the lower isophotes.

To judge the H I extent of "normal" disks, we use the above sample of Wevers (triangles), the sample of field galaxies compiled by Broeils 1994 (squares), and the comparison sample of field spirals of Cayatte *et al.* 1994 (hexagons; excluding galaxies included in the two aforementioned samples). For these samples, we plot the areas subtended at $N_{HI}$ =1.26$\times10^{20}$ cm$^{-2}$ (=1 $M_\odot$ pc$^{-2}$; open symbols; 132 points) and at the last measured H I data point (filled symbols; 99 points). We note that in Fig. 9b, two of the comparison points fall off the plot to higher areas at $N_{HI}$ =1.26$\times10^{20}$ cm$^{-2}$ (to a maximum area of 5,730 $h^{-2}$ kpc$^2$ for NGC 5533, Broeils 1994), and five points fall off at the lower column densities, to a maximum area of 16,300 $h^{-2}$ kpc$^2$ (NGC 2998, Broeils 1994)

The values from the comparison samples represent inclination corrected values, and as such indicate the maximum cross-sectional area available at the quoted isophotes. No such correction has been applied to the merging data due to the extremely perturbed outer morphology. These figures show that the merging systems fall in the upper range of the area distributions for normal galaxies (to the extent

**FIGURE 9:** Area subtended by each system as a function of **(a)** $R$-band surface brightness $\mu_R$, and **(b)** H I column density $N_{\rm HI}$. The size of each system is shown with a solid line and different symbols, as indicated in the legend. The comparison samples are composed of isolated field spirals, and are described in the text. We note that the points for the comparison samples have been corrected for inclination and therefore represent the maximum cross-sectional area. No such correction has been made for the merging systems, which fall in the upper range of both area distributions. The horizontal axis in both panels is on the same scale so that the H I to optical cross-sections can be directly compared at any given surface brightness level. At the faintest level, there is little difference between the subtended areas (Table 6). This is in contrast to difference in linear extents (Table 2; $l_{HI}/l_{opt}$ =1–7.5), which simply means that the tidal features have a small area filling factor.

TABLE 6. Ratio of H I to Optical Areas.

| H I Column Density (cm$^{-2}$) Isophote | $10^{20}$ | $2\times10^{19}$ |
| --- | --- | --- |
| Optical Surface Brightness Isophote | $D_{25}^{(a)}$ | $\mu_R=26.5$ mag arcsec$^{-2}$ |
| Field Sample$^{(b)}$ | 2.4±1.3 | — |
| Merging Sample: | | |
| Arp 295 | 4.4 | 1.2 |
| NGC 4676 | 0.5 | 0.6 |
| NGC 520 | 2.5 | 2.5 |
| NGC 3921 | 1.4 | 1.0 |
| NGC 7252 | 1.1 | 1.1 |

Notes to TABLE 6

$^{(a)}$ Optical Area calculated via $\frac{\pi}{4}D_{25}$, where $D_{25}$ is the inclination corrected optical (blue) diameter, taken from the RC2.

$^{(b)}$ Comparison sample of 84 field spirals taken from Table 4 of Cayatte *et al.* 1994.

that field spirals are "normal"). However, there are still a fair number of "normal" systems that subtend a larger angle than any of the observed mergers: 17% of the comparison sample lie above the Arp 295 curve in the plot of optical cross-sections, while 9% lie above the Arp 295 curve in the plot of H I cross-sections.

Another question of interest is whether the ratio of the H I to optical areas is increased due to mergers. Since the H I disks in normal spirals generally extend beyond the optical disks (Bosma 1978, Cayatte *et al.* 1994, Broeils 1994), and the outer regions obtain higher expansion velocities during the encounter (TT), one might expect this ratio to grow with time.

In an attempt to address this question we use two measures of the H I to optical area cross-sections for the mergers, giving the results in Table 6. In the first column we list the H I area subtended at $N_{HI}=10^{20}$ cm$^{-2}$ divided by $\frac{\pi}{4}D_{25}^2$ for each system. This can be compared with similar measurements of the Cayatte *et al.* (1994) sample, also listed in the table. In the other column we list the H I area measured at $2\times10^{19}$ cm$^{-2}$ divided by the optical area subtended at $\mu_R=26.5$ mag arcsec$^{-2}$ (*i.e.* lowest greyscales and contours in Fig. 4). The field galaxy data are not sensitive enough to derive an equivalent measure.

This table shows that at the faintest level there is little difference in the areas over which the gas and stars are distributed. This is somewhat surprising given the very different optical and gaseous morphologies evident in Fig. 4. Therefore while the increase in linear extent of tidal features is rather impressive (Table 2), the cross-sectional areas *at a given isophote* are not much larger than in normal spirals. Finally we note that there is no trend of cross-sectional size with the stage of merging.

# 6. Discussion

The main issue we wish to discuss is: what are the effects of such major mergers on the progenitors? In particular, what happens to their ISM? We will tackle this issue in two parts, examining the effects on the inner regions in §6.1 and on the outer regions in §6.2. The final issue we examine is what will these systems look like in the next few Gyr? Will these former disks indeed become indistinguishable from modern day ellipticals? And if so, do they offer us any clues as to how to distinguish between ellipticals formed by major mergers and other processes? Are there other fates in store for some of these systems? These questions will be addressed in §6.3.

## 6.1 Effects of Merging on the Inner Regions

### 6.1.1 Starlight

Both NGC 3921 and NGC 7252 already resemble ellipticals in their photometric properties, with $r^{1/4}$ radial light profiles and obeying the Faber-Jackson relationship (*cf.* Schweizer 1995). The gas-rich nature of their tidal tails and the gas kinematics are undeniable evidence that these systems are disk-disk merger remnants. That these systems now lack large stellar or gaseous disks confirms that the merging process can alter the identities of the progenitors, rearranging the majority of the light into a single centrally relaxed $r^{1/4}$ light profile (Schweizer 1982, 1995, Stanford & Bushouse 1991, HGvGS). The $r^{1/4}$ profile is a natural consequence of violent relaxation (Lynden-Bell 1967, van Albada 1982), whereby the rapidly changing potential of the merging bodies scatters stars onto a wide range of orbits (*cf.* Barnes & Hernquist 1995).

HGvGS found that NGC 7252 also fits within the fundamental plane defined by elliptical galaxies (Djorgovski & Davis 1987), showing that at least some disk-disk mergers can get around the objection that ellipticals have higher phase space densities than spirals (Ostriker 1980, Carlberg 1986; see also TT, Negroponte & White 1983, Veeraraghavan & White 1985, Barnes 1988, Lake 1989, Kormendy & Sanders 1992). The tail kinematics, E+A spectrum (Schweizer 1990, Fritze-von Alvensleben & Gerhard 1994), and UBV colors (Schweizer & Seitzer 1992) date these objects at about a Gyr or so since the merger (Schweizer 1995). Therefore the primary photometric characteristics of ellipticals are imprinted upon the remnants rather rapidly in terms of the total merging time-scales (see also Doyon *et al.* 1994).

### 6.1.2 Ionized Hydrogen

In the early systems with a high inner gas mass fraction ($[M_{gas}/M_{dyn}]_{inner} \gtrsim 0.08$ $h^{-1}$), we see vigorous disk-wide star formation along with minor axis plumes of warm gas. These plumes may be examples of the galactic "blow-out" phenomena (*e.g.* Heckman *et al.* 1990), whereby the energy input from massive stars and supernovae (SNe) creates a bubble of hot plasma which expands along the minor axis. The high concentration of gas indicates a dense ISM, which may be required to provide the raw material for the plumes and to inhibit the expansion perpendicular to the minor axis (Suchkov *et al.* 1994). In the later stages (NGC 3921, NGC 7252) the H$\alpha$ emission is mostly confined to the inner kpc, where it is presumably associated with a small star-forming molecular gas disk.

All stages of merging show high rates of central star formation, with central H$\alpha$ luminosities (line 12, Table 3) a factor of twenty over the mean of non-interacting samples (Keel *et al.* 1985, Kennicutt *et al.* 1987, Bushouse 1986). This suggests that such interactions elevate the *central* star formation rate for an extended period. The total *global* $L_{H\alpha+[NII]}$ emission (line 13, Table 3), on the other hand, is not large compared to non-interacting late-type spirals (compare values in Table 4 with Fig. 7 of Kennicutt *et al.* 1987).

In a related vein, these systems have infrared properties ($L_{\rm IR}/L_B$, $L_{\rm IR}/M_{\rm H_2}$; Table 3) that are closer to the values measured for normal spirals than the extreme values found for the ultraluminous IR galaxies, which are believed to be in the throes of merging (Lonsdale *et al.* 1984, Joseph & Wright 1985, Schweizer 1986, Sanders *et al.* 1988b). This corroborates studies which show that a disturbed optical morphology does not ensure a highly elevated level of star formation activity (*e.g.*, Joseph *et al.* 1984, Keel *et al.* 1985, Cutri & McAlary 1985, Bushouse 1986, 1987, Solomon & Sage 1988, Stein 1988, Sulentic 1989), suggesting that interactions act primarily to concentrate modestly enhanced disk-wide star formation to within the inner few kpc (Hummel 1981, Keel *et al.* 1985, Bushouse 1987, Wang & Helou 1992). Further, it suggests that merger-induced super starbursts (Joseph & Wright 1985) are either localized in time (Schweizer 1986) or require specific encounter geometries (Hibbard 1995, Ch. V), gas contents (Sanders *et al.* 1986, Mirabel *et al.* 1990), or progenitor structures (Mihos & Hernquist 1994).

### 6.1.3 Cold Gas

In advanced mergers the molecular and atomic gas phases have very different distributions, with the molecular gas concentrated within the remnant bodies and the atomic gas predominantly in the outer regions. High resolution synthesis mapping of two of these systems (NGC 520 Sanders *et al.* 1988a; NGC 7252 Wang *et al.* 1992) shows that the molecular gas forms a highly concentrated rotating nuclear disk with a column density of order $10^3$ $M_\odot\,{\rm pc}^{-2}$. These disks comprise a significant fraction of the central dynamical mass, with values that are several times higher than observed in normal disk galaxies (Kormendy & Sanders 1992, Downes *et al.* 1993). This indicates that the disk gas has lost much of its angular momentum and energy, enabling it to flow into the center (TT, Negroponte & White 1983, Noguchi 1988, Scoville *et al.* 1989, Barnes & Hernquist 1991, 1995).

However, the same process which leads to a central concentration of molecular gas fails to concentrate a significant quantity of atomic gas in these same central regions. On the contrary, the late stage mergers are decidedly poor in their central atomic gas content (Fig. 8), while the tails are gas-rich (Fig. 7). Because of the symmetric forces experienced during the merger, at most one-half of the outer disk can be drawn into a tail. Therefore, *at least* as much H I was sent into the inner regions of the remnant as now resides within the tails (Hibbard & Mihos 1995). Since less than one-quarter of the total H I is found within these regions, we conclude that most of it must have been converted into some other phase during the merger.

This raises two questions: (1) How and when was the atomic gas so efficiently converted into other phases within the remnant body, and (2) where is this former atomic gas today?

We believe that the answer to the first of these questions is given in Fig. 4, where we see that the atomic gas content decreases in concert with the destruction of the optical disks (§5.4). This suggests that as violent relaxation rearranges the disk light into an $r^{1/4}$ profile (§6.1.1) it simultaneously leads to the near total conversion of atomic gas into other forms within the remnant body. During this process the gas will not remain on closed orbits, and will be compressed as it meets other gas streams. In these converging flows the neutral hydrogen may be ionized in shocks, or dissipate, move inward, and change to molecular form. Conversely, if the merger does *not* completely redistribute the disk material (*e.g.*, NGC 520), the gas may remain on non-intersecting orbits, avoid significant compression, and thereby retain its atomic form.

There are three likely reservoirs for the former atomic gas: (1) stars, (2) molecular gas, and (3) hot gas. The observations of the remnants NGC 3921 and NGC 7252 provide evidence for the presence of all three forms. These are discussed in more detail in Schweizer (1995) and HGvGS, and summarized here. The presence of an E+A spectral energy distribution in each (*i.e.*, the presence of broad Balmer absorption lines; Dressler & Gunn 1983) indicates that a significant portion of gas has been turned into stars (see Schweizer 1995 for NGC 3921 and Fritze-von Alvensleben & Gerhard 1994 for NGC 7252). Additionally, while NGC 3921 and NGC 7252 are atomic gas rich with respect to ellipticals or S0's (see §6.3.1) they have below normal molecular gas contents for their spiral progenitors (Solomon & Sage 1988, Young & Knezek 1989), thus there has been little *net* conversion of H I into molecular form[8]. Finally, the x-ray component detected by ROSAT in NGC 7252 (HGvGS) raises the possibility that some of the atomic gas has been heated to x-ray temperatures, via high velocity cloud collisions (Harwit *et al.* 1987) and/or the energy input from the massive stars and resulting SNe of any nuclear starburst (Graham *et al.* 1984, Chevalier & Clegg 1985, Heckman *et al.* 1990). Unfortunately, the present data do not allow for a detailed accounting between the different phases.

The current hydrodynamical codes nicely support the above results. Although they do not track the gas phase, they do track the density and temperature of the gas. In these simulations (Barnes & Hernquist 1995 and references therein), the majority of the gas becomes concentrated in a dense knot at the bottom of the potential which may reasonably be identified with the central molecular gas concentrations. Any diffuse gas remaining within the main body is virialized (temperatures of order $10^6$ K), aside from streams of cold gas falling back in from the tidal regions. These components would be identified with an x-ray halo and the returning tidal H I.

---

[8] Note: This is a different state of affairs than for on-going mergers which are IR luminous (as opposed to these more evolved merger remnants). Such galaxies also have high ratios of $M_{H_2}/M_{HI}$ (Mirabel & Sanders 1989), but this is due to unusually high values for $M_{H_2}$ (e.g. Sanders *et al.* 1986, Mirabel *et al.* 1990, Sanders *et al.* 1991) rather than a low $M_{HI}$.

### 6.2 Effects of Merging on the Outer Regions

It was two decades ago that Toomre & Toomre (1972) demonstrated so effectively that tails are the natural result of the tides raised during close encounters. Soon after, observations of H$\alpha$ emission in the tails of four of the five systems presented here were measured and found to be consistent with a tidal origin (Arp 295: Stockton 1974b; NGC 4676: Stockton 1974a; NGC 520: Stockton & Bertola 1980; NGC 7252: Schweizer 1978). Much more sophisticated work has been done on this problem since the Toomres 120 massless test particles threaded their way through the orbiting static potentials (see Barnes & Hernquist 1992b and references therein), but this remains a generic result: gravity is the prime mover in the formation of tails. As a result, we observe long tidal features composed of gas and stars exhibiting smooth runs in velocity.

Two observational consequences of the gravitational development of the tails are the possibly self-gravitating clumps of stars, gas, and star formation found within them, and the streams of tidal material falling back into the late stage remnants. Both of these have been discussed in the recent literature, and we briefly summarize the main results below. We then discuss the evolution of the outer tails, and the differences between the gaseous and stellar morphologies of the tidal structures.

#### 6.2.1 Clumps in Tails

The propensity for giant H II regions and clumps of stars to occur at the optical ends of tidal tails was pointed out by Schweizer (1978), who included both NGC 3921 and NGC 7252 as prototypes. The higher H I velocity dispersions at the location of these clumps is possible evidence that they are self-gravitating (see Hibbard & van Gorkom 1993 and HGvGS for velocity dispersion maps). The velocity fields (Fig. 5) continue smoothly through the location both of these clumps, and they therefore lie within the tails.

The observed clumps have luminosities, H I contents, and masses similar to dwarf irregular galaxies (*cf.* Hunter & Gallagher 1985, Freeman 1986). Recent unsuccessful attempts to detect CO in some of these clumps show that they may resemble dwarfs also in their low CO abundance (Smith & Higdon 1994). Duc (1995) has further shown that 14 condensations found within the tidal debris of several interacting/merging systems fall within the same region of the surface brightness–absolute magnitude diagram as dwarf irregulars. The formation of self-gravitating clumps within tidal tails is supported by numerical simulations of mergers (Barnes & Hernquist 1992a, 1995, Elmegreen *et al.* 1993, Hibbard & Mihos 1995)

Such self-gravitating clumps accrete material at the expense of adjacent regions (Hibbard 1995 Append. VIII), and may evolve to resemble detached dwarf satellites in orbit around the remnants, as originally suggested by Zwicky (1956). The model of NGC 7252 predicts that the clump in the NW tail of this system will orbit between 14 and 120 $h^{-1}$ kpc with a period of 4 $h^{-1}$ Gyr (Hibbard & Mihos 1995). If this feature is dynamically bound, it has enough mass to avoid tidal stripping (HGvGS) and should be a long-lived companion of the remnant.

### *6.2.2 Returning Tidal Material*

The kinematics of the NGC 7252 tails led to the prediction (HGvGS) and numerical verification (Hibbard & Mihos 1995) that the material at the base of the tails was not expanding outward, but was instead streaming back into the remnant body. This reflects the fact that the vast majority of tidal material remains bound to the remnant (Barnes 1988, 1992, Hernquist 1992, Barnes & Hernquist 1995). As a result, the more tightly bound tail material (*i.e.* those regions of the tails closest to the remnant) will have shorter periods than the more distant regions and will reach their orbital apocenters and move back inward in the potential while the outer regions of the tail are still expanding. Since energy and angular momentum increase monotonically with distance along the tails, the bases of the tails will fall back quickly to smaller radii, while the more distant regions fall back ever more slowly and to larger radii (Hibbard & Mihos 1995).

Since the tidal material is dynamically cold (*i.e.* it has a relatively small velocity dispersion), it is expected to wrap coherently in the central potential, producing loops, shells, ripples, and other fine structure features (Hernquist & Spergel 1992). The high quantities of fine structure observed in the NGC 3921 and NGC 7252 remnants (Schweizer & Seitzer 1992), coupled with kinematic evidence of material streaming back in from the tidal regions in NGC 7252 and possibly NGC 3921, serves as direct verification of this process. This delayed return of tidal material can stretch the duration of the "merger remnant" stage to many Gyrs (Hibbard & Mihos 1995).

Because of these effects, there will be a rapid luminousity evolution of the tails, with short luminous tails raised initially, yielding to longer, less luminous tails at later stages. This effect will be compounded by the fact that star formation appears to peak within the tails before the disks merge, boosting their luminosity (see also Wallin 1990), and that only low mass stars (M<$2M_\odot$) will survive in old tails ($t$ >1Gyr). These effects will need to be considered when using the appearance of tidal features in distant samples to judge merging rates (see also Mihos 1994).

### *6.2.3 Expanding Tidal Material*

Another aspect of tail development is that since the tidal features are created from luminous disk material drawn out over large distances, these features must necessarily have lower surface brightnesses and column densities than the preencounter disks. This may explain why the cross-sectional areas of the mergers are not necessarily larger than isolated disks at a given isophote (Fig. 9). As a result, both the light and gas may have even more extended features at even lower column densities than plotted in Fig. 9.

However, the H I profiles do not seem to extend uniformly to lower column densities. In fact, the lowest observed column densities are similar to those found in normal spirals. In all cases there is a change in the slope of the column density profiles around $N_{\rm HI} \sim 5 \times 10^{19}\,{\rm cm}^{-2}$ (and at $10^{20}\,{\rm cm}^{-2}$ for NGC 4676). A similar cut-off in H I column densities is found in isolated spirals (van Gorkom 1993), and may be due to the ionization of the lower column density H I by an intergalactic radiation field (Maloney 1993) or external pressure confinement by a hot intergalactic medium (IGM).

While the majority of the tidal material remains bound to the remnant, some fraction of it will have enough energy and angular momentum to remain far from the remnant for very long times, forming far-reaching tails stretching over hundreds of kpc in radius. For example, in the NGC 7252 simulation, 20% of the tidal material remains beyond 70 $h^{-1}$ kpc for over a Hubble time. This material would be far too faint to observe in emission, but ought to contribute material to the Ly$\alpha$ forest (Carilli & van Gorkom 1992; Morris & van den Bergh 1994) and to the IGM.

### 6.2.4 Differences Between Gas and Light Distributions

One of the most intriguing properties of these systems is the extensive, low surface brightness purely optical or gaseous features: the outer gaseous ring in NGC 520, the extended halos of light in NGC 4676 and NGC 7252, the plumes of light in NGC 520 and NGC 3921. We find that dividing the outer features into three different classes (tails, plumes and halos) separates them according to their local gas content (§5.5). Since gravity itself does not differentiate between gas and stars, other factors must be at work. Non-gravitational effects which may be playing a significant role are the different velocity dispersions of the stars and gas, the finite ages of stars, the ability of the gas to dissipate, differences in the gas content of the progenitors, or external forces such as an IGM. We examine possible explanations for the different features in turn.

### (i) Gas rich tails and their different optical and gaseous extents:

Fig. 7 shows that $M_{\rm HI}/L_R$ for the late stage mergers is higher than in either the early stage mergers or the outer regions of the field spirals. This may be a natural consequence of the fact that the velocity dispersion of the old stars is higher than the gas (Mihalas & Binney 1981), causing them to disperse over a larger area and leading to an increase in $M_{\rm HI}/L_R$ with time. Self-gravity and gaseous dissipation may also be playing a role, causing the gas to bead along a ridge line in the tail (e.g., Wallin 1990, Barnes & Hernquist 1995). The difference between the velocity dispersions of the tail components would also explain why the tails have faint halos of light (see Figs. 2&4, especially the relationship between the two lowest light contours in Fig. 4): the old stars will make higher excursions both perpendicular and parallel to the tails than either the young stars or the gas.

This same explanation, combined with projection effects, may also explain why the optical tails in Arp 295 and NGC 4676 extend beyond their gaseous counterparts by about the same factor as the faint halos extend beyond the brighter spines of the tails. We fully expect that $M_{\rm HI}/L_R$ increases with distance along the tail, just as it increases with radius in field spirals (Bosma 1978, Wevers *et al.* 1986). However, the outer optical contours of Arp 295a and NGC 4676a in Fig. 4 do not necessarily delineate the "ends" of the tails. Both of these tails are seen nearly edge-on, and the outer optical contours are similar to how the tail of NGC 4676b would appear if it were to be viewed perpendicular to the plane of the sky (*i.e.* from the left-hand side of the page in Fig. 4b). In this tail, the gas-rich extension to the west of the end of the tail suggests that the tail bends away from our line of sight at this point. If viewed from the side, the gas-poor halo of light extending directly south in NGC 4676b and would appear as an optical extension similar to those seen in Arp 295a and NGC 4676a.

With time, these halos will become hopelessly faint, as the stars spread out over a much larger area than the H I (which may also be limited in its extent by the ionization or external confinement of the H I at the lowest column densities, *cf.* §6.2.3). Additionally, the low dispersion stellar disk component composed of young stars will begin to die as the tails grow, leading to a further increase in $M_{\rm HI}/L_R$ with time. Purely gaseous extensions to the optical light are the natural result of these effects, such as those observed in NGC 3921 and NGC 7252.

This explanation has two observable consequences: (1) the gas-poor halos of the tails should be redder than their gas-rich spines; and (2) there should be a broader, much fainter optical counterpart to the gaseous extensions.

### (ii) Gas-poor halos

The dissipative nature of the gas may explain the broad gas-poor halos of light surrounding the remnants. When galaxies merge, they do so with a large "splash", as violent relaxation scatters material onto more radial orbits (§6.1.1). These radial orbits pass through the dense inner regions repeatedly, intersecting many other plunging orbits. Due to its dissipative nature, the atomic gas cannot pass through the central regions unscathed, as it would either shock heat or cool and condense (*e.g.*, Weil & Hernquist 1993, Barnes & Hernquist 1995).

### (iii) Gas-poor and gas-rich plumes

There are two characteristics of plumes which need to be accounted for: their gas content and their plume-like (as opposed to tail-like) appearance.

The plume to the east of Arp 295b is gas rich ($M_{\rm HI}/L_R \gtrsim 2\ M_\odot L_\odot^{-1}$), indicating a gas-rich progenitor. Here, we suspect that the plume-like nature of the light is due to the retrograde encounter. Such encounters fail to raise well defined tails, provoking a broader, milder response in the hosts disk (TT, White 1979). As such, it will remain primarily close to and in rotation around Arp 295b. Upon the next encounter with Arp 295a, it will either be destroyed or settle into a disk around the remnant (*à la* NGC 520). We therefore expect that after the final merger it will be either a gas-poor halo or a gas rich disk, and not a gas-rich plume.

The extensive plumes in NGC 520 and NGC 3921 are gas-poor ($M_{\rm HI}/L_R \lesssim 0.1$). If these features were simply tails viewed face-on, we would expect associated H I column densities of $\gtrsim 10^{20}\,{\rm cm}^{-2}$. We suggest instead that these features arose from progenitors with gas poor disks, *i.e.* S0 to Sa galaxies (see Appendix, §A-3 & §A-4). Such disks lack not only significant amounts of gas, but also a significant population of young stars, and the lack of edge-brightened features in the plumes could be due to the lack of this low velocity dispersion component. This may also explain why plumes as a class have redder colors than tails (Schombert *et al.* 1990).

### *(iv) Gaps in tails*

Gaps in both gas and light may be the natural consequence of self-gravity within the tails, as bound regions accrete nearby material at the expense of the adjoining regions. Star formation may cause local decreases in $M_{\rm HI}/L_R$, as $L_R$ rises due to the light of young stars and $M_{\rm HI}$ drops as gas is consumed by star formation and ionized or swept out due to the energy input from SNe and high mass stars. Evidence for in situ star formation is seen in abundance in tidal tails (Fig. 3; see also Schweizer 1978, Schombert *et al.* 1990, HGvGS, Hibbard 1995). The patchy optical appearance of the tails fills in on deeper exposures (Figs. 1&2), and may reflect an age spread in the underlying stellar component. This delicate interplay between the disruptive forces of star formation and the attractive force of self-gravity may make the tails lively arenas for eons to come.

### *(v) Displacements between optical and H I maxima*

Significant displacements are observed in the Arp 295, NGC 520, and NGC 3921 systems. In the Arp 295 system (Fig. A-1a), the morphology suggests that the stars and gas are both part of the same feature (*i.e.* the bridge or tail), simply with their peaks displaced. An even more pronounced example of this effect is seen in the NGC 4725/47 system (Wevers *et al.* 1984). The reader is referred to that work for a discussion of two possible explanations: angular momentum redistribution due to collisions between gas clouds, and interaction of the tail gas with a dense IGM. Another possibility is that the atomic gas is consumed, ionized, or mechanically displaced from the optical peaks due to in situ star formation. The tenability of the first suggestion is supported by recent hydrodynamical numerical simulations, which show a striking separation between the stellar and gaseous components along a bridge, very similar to that seen in Arp 295, (Mihos, personal communication); however no such separations occur in the tails.

The displacements in NGC 520 and NGC 3921 (Fig. A-1c,d) appear different in nature from those in Arp 295. In these systems, we believe the gas to be from one of the progenitors and the offset stars to be from the other system which is gas-poor (see Appendix). The morphology of these features suggests that there is some interaction between the gas and stars, although we are at a loss as to what this interaction is.

### 6.3 FUTURE EVOLUTION

Both of the late-stage mergers in this study, NGC 3921 and NGC 7252, have already successfully evolved into S0s (Table 3), and have been imprinted with the photometric properties of ellipticals (§6.1.1). In this section we ask what evolution is either likely or required for these remnants to fully evolve into ellipticals, and how they might appear in the interim. This in turn may guide us to judge better the age of such remnants.

#### *6.3.1 From Spiral to Elliptical*

In a recent study, Schweizer (1995) showed that as far as luminosity, brightness distribution, color gradients, velocity dispersion, UBV colors, and fine structure are concerned, NGC 3921 and NGC 7252 will closely resemble normal elliptical galaxies within a few billion years. The most pertinent question here is whether the same can be said for their gas properties?

While the merging process can convert most of the inner H I into other forms before the tidal features fade away (§6.1.3), there are still significant amounts of cold gas associated with these remnants — within the atomic gas-rich tidal features (most of which we expect to turn around and move back inward; §6.2.2), and in small ($r < 1\ h^{-1}$ kpc) star-forming circumnuclear molecular gas disks (Wang *et al.* 1992, Yun & Hibbard in preparation). As a result, they are quite gas rich when compared to E's and S0's (Knapp *et al.* 1985, Lees *et al.* 1991, Bregman *et al.* 1992; see Dupraz *et al.* 1990 for a discussion on NGC 7252). To be truly representative of early type galaxies, NGC 3921 and NGC 7252 need to rid themselves of about half of their present molecular hydrogen supply and 90% of the returning atomic hydrogen.

We believe that this is likely for the following reasons:

- There is a modest amount of star formation taking place in the central regions of both systems (1–2 $M_\odot$ yr$^{-1}$; Schweizer 1995, HGvGS). At these rates, the remaining molecular gas should be turned into stars within the next few Gyr (Dupraz *et al.* 1990, Whitmore *et al.* 1993).

- In NGC 7252, some process is efficiently converting the returning H I into other forms (HGvGS). The NGC 7252 simulation suggests that half of the present tidal H I will fall to within ∼13 $h^{-1}$ kpc of the remnant body in the next 2.4 $h^{-1}$ Gyr. The gas at the base of the NW tail is seen at a projected radius of about 10 $h^{-1}$ kpc, and the simulation geometry puts it at a physical distance of over twice this (22 $h^{-1}$ kpc). Since this material is currently being converted into other forms at a rate of 1–2 $M_\odot$ yr$^{-1}$, it is possible that much of the remaining H I will also be converted into other forms upon its return.

So while the photometric properties of ellipticals are imprinted rather rapidly upon the remnants (§6.1.1), it will take several Gyr longer before they obtain more representative quantities of cold gas.

### 6.3.2 Long Lived Signatures in Merger Remnants

In the previous sub-section we discussed the likely evolutionary paths which will lead the late stage remnants to resemble ellipticals. In this section we ask the equally important question, *"How and for how long might one distinguish an elliptical made from a disk-disk merger from a truly old elliptical?"*. Since *any* galaxy formation process will involve the same physical processes involved during a merger (*e.g.*, violent relaxation, gaseous dissipation, and star formation), the signatures will not be unique.

The remnants have significant circumnuclear molecular gas disks which will be turned into circumnuclear stellar disks within a few Gyr (§6.3.1). HST has imaged the stellar disk associated with the molecular gas disk in NGC 7252 (Whitmore *et al.* 1993). The kinematic signatures of such stellar disks are found with increasing frequency in the centers of otherwise normal looking ellipticals (*cf.* Bender 1990 and references therein; Bender & Surma 1995), and are a possible signature of a merger origin.

There is approximately 15–20% of the total $R$-band light found in bright loops and shells surrounding the remnants, and gas-rich tidal material will continue to fall back towards to remnant, at ever decreasing rates and to ever larger radii (§6.2.2). The stellar tails and loops will continue to phase wrap coherently in the central potential, producing yet more fine structure features (Hernquist & Spergel 1992). With time, these structures will become increasingly difficult to uncover, as they lose phase coherence with each radial orbit, and as the surface brightness falls off with radius (Hibbard 1995, Append. VIII). Therefore, the fraction of light in such non-axisymmetric features should decrease at a rate proportional to the crossing time of the remnant.

The H I streaming back in from the tails is presently converted into other forms (§6.3.1). If this material is being shock heated or evaporated by a recently formed hot halo (§6.1.3), we expect the x-ray luminosity of the halo to grow as the remaining H I and CO contents decrease. This is an attractive possibility, as it would naturally account for the fact that early-type galaxies have more hot gas than late types of the same blue-luminosity (Fabbiano *et al.* 1992) and have a hot gas content which is anti-correlated with their cold gas content (Bregman *et al.* 1992). If instead the atomic material it is being compressed into molecular form, we have a steady state between the gas accretion rate and the present star formation rate and the system will remain molecular gas rich for a very long time ($\gtrsim$5 Gyr).

At some point, the late returning tidal H I will not approach the remnant closely enough to be converted into other phases. We would then expect to see some outer atomic hydrogen around old remnants. Due to orbit crowding upon its return, the gas will have the highest column densities near its pericentric distance. This neutral gas should appear as incomplete rings at large radii primarily in rotation (since the radial velocities near turn-around are small). Such partial rotating rings of gas are being found frequently in VLA observations of shell galaxies (Schiminovich *et al.* 1994, 1995).

Meanwhile, the outer tidal regions will continue to expand (§6.2.3). In the absence of external forces (*e.g.*, neighbors, IGM, cluster potential), these expanding regions will lead to large, low column density extensions of faint light and gas around old mergers. Within a few Gyr, we would expect the tails to take on a patchy appearance, with partial optical and gaseous tidal filaments, and perhaps mostly detached satellite dwarf companions assembled from the former outer disk material (§6.2.1).

Therefore, we expect a disk-disk merger remnant to reveal its merger origin by extended H I and optical tidal filaments, an excess of molecular gas and optical fine structure features, an aging post-burst population, and a small star forming circumnuclear gas disk, all of which should persist for several Gyr. As these measures fade, we expect them to be replaced by a growing circumnuclear stellar disk and perhaps an x-ray halo, partial rings of H I at large radii, partial tidal filaments possibly pointing to youngish dwarf companions, and envelopes of very extended gas and stars at very low surface brightness levels. Any one of these characteristics alone would not unambiguously mark a remnant as the product of a major merger, but the presence of most would seem to favor such an origin over, for instance, many smaller accretion events.

### *6.3.3 Other Fates*

It is not obvious that all of the present merging systems will end up as ellipticals or lenticulars. NGC 520 strikes us in particular as a system with a very different fate in store. There is a sizable inner disk extending to 20 $h^{-1}$ kpc, and most of the outer H I is primarily in rotation. The optical and infrared isophotes suggest that the merging nuclei are embedded well within this inner disk (see §A-3). Since it has survived the merger up to this point, it is not obvious that the final coalescence will have much impact on the gas disk.

At present, the H I column densities in the outer regions are lower than the threshold for star formation (Kennicutt 1989), but the gas may settle in the future and obtain such densities, laying down a "young" disk of stars. In this case, in a few Gyr, would this system not be a spiral?

The factors which left NGC 520 with its gas disk intact remain hidden to us. Numerical simulations suggest that the mass ratio is of critical importance, as the dynamically cold disks are easily disturbed by the violently fluctuating fields that accompany high mass-ratio interactions ($M_1/M_2 >0.3$) (Quinn *et al.* 1993). However, these simulations were purely stellar-dynamical, and they should be repeated with a significant dissipative (gaseous) component. It is likely that encounter geometry also plays a significant role, and certainly one of the progenitors had to have had a massive gaseous disk.

Therefore, the gas-poor nature of the remnant may depend on factors other than just merging, and it is possible that a wide range of galaxy types — S0's, dusty ellipticals, and even spirals — may be the end product of such mergers. We conclude that while the evidence is very strong that some mergers will evolve into ellipticals, mergers do not necessarily *have* to make ellipticals. Much more work lies ahead before we can predict which types of mergers will make which types of galaxies.

## 7. Summary & Conclusions

In this paper we have presented spectral line H I mapping and deep $R$-band and H$\alpha$ imaging of five disk-disk mergers in progressively advanced stages of merging. The following are the main conclusions that we draw from these data:

- Moving along the sequence from early to late-stage mergers, a larger fraction of the HI appears outside the optical bodies, in extended tidal features. In the final stage all of the H I is found in the tails, with no atomic gas detected within the remnant body. We suggest that as the process of violent relaxation redistributes the original disk stars into a relaxed profile, it causes the majority of atomic gas within these disks to compress, condense, form stars, and/or shock heat to x-ray temperatures, with any residual atomic gas in the remnant heated by the energy input of the resulting starburst.

- The two merger remnants, NGC 3921 and NGC 7252, already have the photometric properties characteristic of ellipticals, and will likely evolve to even more closely resemble ellipticals in a the future (Schweizer 1995). Our observations show that these systems are likely to resemble ellipticals also in their cold gas contents. Therefore the idea that at least some spiral galaxies can merge and evolve into ellipticals seems to be well established.

- The observations of NGC 520 reveal an extensive rotating disk of $4\times10^9\,h^{-2}\,M_\odot$ of H I centered on the primary nucleus. Since the optical isophotes suggest that the two galactic nuclei are embedded within this disk, we conclude that mergers may not necessarily destroy the gaseous disks of the progenitors and therefore need not evolve into ellipticals.

- The H$\alpha$ imaging shows that three of the five galaxies in the early and intermediate stages of merging exhibit plumes of ionized gas emanating from the nuclear regions along the minor axis, suggesting that they are experiencing galactic-scale "blow-outs". These plumes appear in the galaxies with the highest central gas mass fractions.

- The tail kinematics are determined mostly by gravity. Star formation occurs within the tidal features at all stages. At the later stages there are star-forming clumps of stars and gas which may be self-gravitating and have global properties similar to dwarf irregular galaxies. These systems should evolve into dwarf satellites around the remnants.

- In the late stages of merging the bases of the tidal tails are seen falling back towards the remnant body, while the ends are still kinematically expanding. The tidal material which remains bound to the remnants will continue to rain back upon it for several more Gyr, in ever decreasing amounts and to ever larger radii, while the unbound material will lead to very faint extensions of light and gas. That the remnant bodies remain H I poor suggests that the returning tidal material is continuously converted into other forms upon its return.

- There are some subtle but important difference in the distribution between the light and gas, especially at the fainter levels. These range from gaps and displacements to extensive optical/gaseous features without gaseous/optical counterparts. Some of the observations may be due to the nature of the progenitors (*e.g.* gas poor disks), and some due to kinematic differences between gas and stars (*e.g.* dissipation, velocity dispersion). These differences need to be explored explicitly with detailed models.

- The presence of atomic gas at large radii, circumnuclear molecular and stellar disks, optical "fine structure", an aging central post-burst stellar population, and perhaps an x-ray halo may be the best clues for identifying more evolved merger remnants. The amount of atomic and molecular gas and optical fine structure should decrease uniformly with time, while any stellar disk and/or x-ray halo should grow. Since tidal H I is the dynamically coldest and most loosely bound component of the progenitors, it should leave the longest-lived features at large radii.

This work supports previous work that suggests that mergers can indeed transform gas-rich disk systems into (relatively) gas-poor dynamically relaxed systems. However, there are many possible progenitors and encounter geometries, and we have only observationally tracked mergers that lead to the formation of long tidal tails. Multiwavelength investigations like the present one should help to reveal the observational signatures of different kinds of mergers, so that we can move beyond the question of whether galaxies merge to answer the question "the merger of what?". This is a necessary step to determine the merger frequency rate and answer the ultimate question of how and what type of galaxies are born.

We would like to thank François Schweizer for the original suggestion for this project, and particularly for his help in selecting the objects for study. We are also indebted to him for his many careful re-reading of the manuscript and his patience and undaunted enthusiasm over the years it took to bring it all together. We also thank Raja Guhathakurta, David Helfand, Michael Rich and Dave Schiminovich for a lively discussion and their constructive input on an earlier version of this paper. JEH thanks Dave Westpfal for introducing him to the finer points of optical imaging, Raja Guhathakurta for help with the mosaicing technique, and Josh Barnes, Chris Mihos, Rob Olling and Hong Sheng Zhao for many helpful discussions.

We would like to thank Françoise Combes, Jason Surace, Chris Mihos, and James Schombert for the sharing of unpublished data; Adrick Broeils and Veronique Cayatte for supplying their comparison sample data; and Theresa McBride for help with the color images. JEH offers special thanks to NRAO and NOAO for the generous allocations of observing time for this project, and the VLA staff for their time, comments, hospitality, and indulgence over the last 5 years.

This work has been supported in part by NSF grants AST 89-17744 and AST 90-23254 to Columbia University, and by Grant HF–1059.01–94A from the Space Telescope Science Institute, which is operated by the Association of Universities for Research in Astronomy, Inc., under NASA contract NAS5–26555. This research has made use of the NASA/IPAC Extragalactic Database (NED), which is operated by the Jet Propulsion Laboratory, California Institute of Technology, under contract with the National Aeronautics and Space Administration.

<div align="center">References</div>

46                                HI, HII, and R-band Observations of a Merger Sequence

# Appendix A: Detailed Description of Individual Systems

In this appendix we present a detailed description and supplementary figures of the H I observations for four of the object in this study: Arp 295, NGC 4676, NGC 520, and NGC 3921. The data for NGC 7252 are fully presented and discussed in HGvGS. In Fig. A-1 we present a montage showing a close-up of a specific region in each of the systems. This montage illustrates differences between the gas and light distributions, and will be referred to in the individual subsections. In Table A-1 we present a list of the H I properties of all systems detected within VLA primary beam (30′). The data presented in each subsection consist of the velocity channel maps and zeroth and first moment maps. A description of the observations and data reduction is given in the main text (§3).

## A.1 Arp 295

As mentioned in the introduction, Arp 295 (=VV 34) was one of the original systems modeled in the "bridge and tail building" paper of TT. It is the most impressive bridge-tail system in the Arp atlas, with the bridge spanning more than 95 $h^{-1}$ kpc across the sky. While an early study attributed the origin of this bridge to entrainment of particles in a magnetic field (Arp 1962), ionized gas rotation curves for the two galaxies (Stockton 1974b) and color photometry of the bridge (Beaver *et al.* 1974) vindicate the Toomre's tidal interpretation. In this model, both the bridge and tail are spun off of Arp 295a during a prograde encounter (see Fig. 19 of TT; since no particles were put in orbit around Arp 295b, its orientation was not constrained). The bridge emerges from the southern edge of the Arp 295a disk and reaches up to Arp 295b, while the southern tail emerges from the northern disk edge, crosses in front of the galaxy, and curves away from us to the south. As will be discussed presently, the H I kinematics support this geometry.

Arp 295a&b are the largest spirals in a loose group of galaxies. Nine small spirals or dwarf irregulars fall within the primary beam (600 $h^{-1}$ kpc) and velocity range (6300–7400 km s$^{-1}$) of the H I observations, with H I masses ranging from 0.3–16×10$^8$ $h^{-2}$ $M_\odot$ (Table A-1). Eight of these are shown in the H I moment map of this system (Fig. A-2), and the ninth lies 21′ to the southeast. Three of the companion systems appear in Figs. 1–5 (G2, G3, G4, and G6 of Table A-1). Two bright ellipticals are also in this field, but since no H I is detected from either, it is not known if they are associated with the group. One of these, IC 1505, appears in the top right corner of Figs. 1–4.

The H II emission is contained mostly within the disks of both galaxies. The only H II emission associated with the tidal features are two unresolved clumps at the base of the southern tail (Fig. 6b), each with a luminosity of order 5×10$^{38}$ $h^{-2}$ erg s$^{-1}$. In Arp 295b there are large plumes or loops of H$\alpha$ emission reaching 6.4 $h^{-1}$ kpc both north and south along the minor axis (Fig. 6a). Arp 295a is seven times fainter in H$\alpha$, although dust in the disk must be blocking some emission. The high inclination of this galaxy is most favorable for viewing minor axis emission, yet none is detected (Fig. 6b). In §5.3 we suggest that the presence of an ionized plume in Arp 295b and not Arp 295a might be related to the fact that the former system has an inner gas fraction (gas mass to dynamical mass; line 8 of Table 4) that is five times that of its southern neighbor ($M_{gas}/M_{dyn} = 0.11$ vs. 0.02).

The velocity field (Fig. 5a, Fig. A-2) indicates that the two disks rotate in the opposite sense: the projection of the angular momentum vector of Arp 295a in the sky plane points to the southeast and that of Arp 295b to the north. This agrees with the ionized gas rotation curves of Stockton (1974b). The rotation curves give a dynamical mass ratio of 3:1 for Arp 295a:Arp 295b. Since Arp 295a is blue-shifted with respect to Arp 295b, we deduce that Arp 295a experiences the encounter in a prograde sense, while Arp 295b is undergoing a retrograde encounter.

The geometry of TT is supported by the H I kinematics as represented by the intensity weighted velocity fields (Fig. 5a and Fig. A-2) and the H I channel maps (Fig. A-3). In particular, the smooth connection in velocity space between Arp 295b and the bridge (at the red and orange colors in Fig. 5a and in the channels between 6972–7014 $\mathrm{km\,s^{-1}}$ in Fig. A-3) suggests that Arp 295b is actually physically accreting material from the bridge, and that the proximity is not a projection effect. The continuity in velocity along the whole length of the bridge (from yellow to green in Fig. 5a; channels between 6847–7014 $\mathrm{km\,s^{-1}}$), and the discontinuity between these velocities and the gas in the northern part of the disk of Arp 295a (between green and red in Fig. 5a) supports the interpretation that the bridge emerges from the far side of Arp 295a and is swinging away from us (TT). Similarly, the smooth decrease in velocity from the southern side of the Arp 295a disk along the length of the tail (channels from 6553–6763 $\mathrm{km\,s^{-1}}$) suggests that the southern tail emerges from the near side and swings towards us.

This system is very rich in both atomic and molecular gas ($M_{gas} = 2.3 \times 10^{10}\, h^{-2}\, M_\odot$). The outer regions contain most of the atomic gas (∼60%), primarily in the optical plume to the east of Arp 295b. This plume has an H I mass to light ratio which is twice as high as in the inner regions of Arp 295b, and 3–5 times higher than found in the bridge and tail (Fig. 7a). This is the only gas-rich plume-like optical feature observed in the present sample. The gas in Arp 295a is confined to the disk and tail, while there is an additional diffuse halo component to the starlight (Fig. 2a and Fig. 4a).

Although the gaseous tidal features bear a general resemblance to their optical counterparts, there are some peculiar differences. The gas on the northern portion of the bridge near Arp 295b is very well aligned with the optical bridge, but becomes progressively displaced to the northwest as the bridge traces back towards Arp 295a (see Fig. 4a). At the region where the optical bridge is projected onto Arp 295a, the gas is displaced by as much as 20″ (6.6 $h^{-1}$ kpc). Along the southern tail we again find systematic displacements between the light and the gas peaks, this time with the gas displaced to the south of the tail (Fig. A-1a). The largest displacement (8″ ∼2.5$h^{-1}$ kpc) is observed at the base of the tail. The proper motion inferred from the above geometry is such that the bridge moves across the sky from northwest to southeast, while the tail crosses from the southeast to the northwest. This suggests that the displaced gas lies on the far side of the stars in terms of its orbital motion.

As a final puzzle, we find that the optical tail of Arp 295a extends 5 $h^{-1}$ kpc beyond the end of the gas distribution (Fig. A-1a). The behavior of the mass-to-light ratios shown in Fig. 7a shows that this is not due to a projection effect, since such projections would lead to similar decreases in the column densities of both components.

## A.2 NGC 4676

NGC 4676 (="The Mice"=Arp 242=VV 224) was also one of the four systems modeled by TT, and again subsequent kinematic observations vindicated the tidal interpretation (Stockton 1974b, Mihos *et al.* 1993). In the model, both galaxies experience a prograde encounter, with their northern edges receding from us and the northernmost galaxy (NGC 4676a) red-shifted with respect to the southern (NGC 4676b). With this orientation, the northern tail is on the far side of NGC 4676a and swinging away from us. NGC 4676b is rotating clockwise and is oriented such that the northeastern portion of the disk is closer to us.

These kinematics are in general agreement with the H I velocity field (Fig. 5b and Fig. A-4) and channel maps (Fig. A-5), with a few notable exceptions. First, the H I rotation curve for NGC 4676a is significantly lower than the H$\alpha$ rotation curve (270 km s$^{-1}$ *vs.* 380 km s$^{-1}$; Table 3). This is likely due to beam smearing of the H I data, since the optical rotation curves turn over within 10″ (Stockton 1974b, Mihos *et al.* 1993), which is half the size of the H I beam. From Fig. A-5 we find emission from the central regions of NGC 4676a over the velocity range 6578–6966 km s$^{-1}$, consistent with the optical velocities. Secondly, we find that the H I velocity width for NGC 4676b is significantly higher than the optical velocity width (420 km s$^{-1}$ *vs.* 270 km s$^{-1}$). The H I isovelocity contours of this region suggest that this is because the kinematic axis of NGC 4676b is offset from the position angle of the bar (position angle of 20°  *vs.* 28°), along which the ionized gas velocities were measured. In the higher velocity resolution C-array data ($\Delta v$=22 km s$^{-1}$) we see H I emission associated with the disk of the southern galaxy over the velocity range from 6277–6751 km s$^{-1}$ (see also Fig. A-5), and suspect that the H I measurements are a more accurate measure of the total velocity range. If this range indicates the width of the rotation curve we derive a mass ratio of 2:1 for NGC 4676b:NGC 4676a.

This system has the same optical luminosity as Arp 295, but less than half of the total gas content ($M_{gas} = 9.8 \times 10^9 \, h^{-2} \, M_\odot$). H I emission is also detected from a small galaxy at the same velocity lying 270 $h^{-1}$ kpc to the south (Table A-1). The tails are quite luminous, containing one-third of the total *R*-band luminosity of this system, with $L_R$= 5×10$^9 \, h^{-2} \, L_\odot$ in the high surface brightness ($\mu_R \lesssim 22$ mag arcsec$^{-2}$) northern tail, and $L_R = 2 \times 10^9 \, h^{-2} \, L_\odot$ in the fainter southern tail. The tails also have a high atomic gas content, with on average $M_{\rm HI}/L_R \sim 0.4 \, M_\odot L_\odot^{-1}$ but going up to 2 at the end of the southern tail (Fig. 7a).

Both disks have widespread H II emission, and within the main body of NGC 4676a there is a plume of H$\alpha$ emission extending 6.6 $h^{-1}$ kpc along the minor axis (Fig. 6c). A similar plume would be difficult to identify in NGC 4676b due to its low inclination. As was the case for Arp 295, the galaxy with the minor axis plume, NGC 4676a, also has the highest central gas fraction ($M_{gas}/M_{dyn} = 0.08$ *vs.* 0.03). NGC 4676b possesses an ionized gas bar, which is offset in position angle by several degrees from the underlying stellar bar. A similar offset bar is produced in hydrodynamical simulations of mergers (Barnes & Hernquist 1991, 1995), in which the angular momentum transfer between the two bars serves to drive large amounts of gas to the center of the galaxy.

The tails have widespread star formation which accounts for 16% of the total H$\alpha$ emission. There are many large knots in the northern tail, with diameters of $\sim$2 $h^{-1}$ kpc and luminosities of order 1–5$\times 10^{39}\,h^{-2}\,\mathrm{erg\,s^{-1}}$, but due to the edge-on viewing geometry it is not possible to say whether these regions are giant H II complexes or smaller complexes projected close to each other. For the southern tail, which has a more favorable viewing angle, we can discern many smaller knots of star formation with luminosities of order $5 \times 10^{38}\,h^{-2}\,\mathrm{erg\,s^{-1}}$. These clumps are unresolved in the 2″ seeing, and therefore have diameters of less than 0.7 $h^{-1}$ kpc. Since the total luminosity of the southern tail ($1.2 \times 10^{40}\,h^{-2}\,\mathrm{erg\,s^{-1}}$) is of the same order as the northern tail ($2.1 \times 10^{40}\,h^{-2}\,\mathrm{erg\,s^{-1}}$), we suspect that much of the northern tail is composed of similar clumps.

While there is a general resemblance between the optical and gaseous tidal morphology, there are again some interesting differences. As in Arp 295a, the stellar tails are more extensive than the gaseous tails (see Table 5 and Fig. 7a), particularly for NGC 4676a (Fig. A-1b). In these extensions, the light drops in intensity by a factor of two while the gas intensity falls by more than a factor of ten over the same distance. Additionally, there is a gap in the gas distribution along the southern tail (see Fig. 3b and Fig. 7a). Since there is only a 20% drop in the light brightness at this point and a factor of five drop in the gas intensity, we conclude that this is a real gap in the gas distribution and not a line-of-sight integration effect. Finally, we find that the faint, broad light distribution seen in Fig. 2b has no gaseous counterpart. Instead, the gas is confined to the spine of the tidal tails (Fig. 4b) while the light has an additional more diffuse component (*i.e.* a halo). Along the spine of the tail there are no significant displacements between the gas and light peaks, although there is more gas to the west of the southern tail (*i.e.*, behind the tail in terms of its proper motion) than to the east.

### A.3 NGC 520 & UGC 957

As the "second brightest very disturbed galaxy in the sky" (Arp 1987) and a charter member of the Irr II class of galaxies (Sandage 1961), NGC 520 (=Arp 157=VV 231) has received much attention in the literature throughout the years. While early investigators believed this system to be a single galaxy undergoing the trauma of an exploding core (Sandage 1961, Arp 1967, Tovmasyan 1967, Khachikyan 1973, Tovmasyan & Sramek 1976), more recent investigations overwhelmingly favor a merger genesis for this system (Vorontsov-Vel'yaminov & Arkhipova 1963, TT, Krienke 1975, Stockton & Bertola 1980, Sanders *et al.* 1988a, Stanford & Balcells 1990, 1991, Stanford 1990, 1991, Bernlöhr 1993).

Spectroscopic and photometric studies support the tidal interpretation and have identified two light concentrations in near IR imagery as the remnant nuclei (Bushouse and Werner 1990, Stanford and Balcells 1990). The primary nucleus is coincident with the center of the dust lane, and the secondary nucleus lies 40″ (4.4 $h^{-1}$ kpc) to the northwest. The locations of these two peaks are indicated by crosses in Fig. A-1c. The strongest evidence that these $K$-band peaks represent true mass concentrations is the kinematics of stellar absorption line spectra which show the velocity dispersions to have marked increases as they cross either nucleus, and near IR photometry which shows the NW nucleus to have the infrared colors typical of normal bulge stellar populations (Stanford & Balcells 1990). The velocity field is continuous between the two peaks (Stockton and Bertola 1980, Stanford and Balcells 1990, Bernöhr 1993), suggesting that they are physically close, and the fainter light appears as a common envelope surrounding the embedded nuclei, rather than the superposition of two light profiles.

Due to the advanced nature of this merger, it is not possible to assign a mass ratio to the progenitors. The primary nucleus is four times brighter than the NW nucleus at 2.2$\mu$m (Stanford & Balcells 1990), but since it is experiencing an ongoing star burst while the NW nucleus is in a post-burst stage (Stanford 1991, Bernlöhr 1993) this ratio will not reflect the underlying mass ratios. Rotation curves give a ratio of 10:1 (Bernlöhr 1993), but the disturbed kinematics between the two nuclei make this determination very uncertain. $^{12}$CO(1→0) linewidths (Sanders *et al.* 1988a) and stellar velocity dispersion measurements (Stanford & Balcells 1990) suggest that the remnant nuclei have similar masses, which might suggest a progenitor bulge mass ratio near 1:1. However, these ratios may not be a good indicator of the mass ratios of the disks.

Our H$\alpha$ image agrees with those published in Bernlöhr (1993) and Young *et al.* (1988), though the present observations reveal quite a bit more structure. In particular, Fig. 6d shows that along with H$\alpha$ concentrations associated with both nuclei and several knots of star formation within the disk and southern tail of NGC 520, there are again minor-axis plumes of ionized gas emission. The southern emission "plume" might owe its morphology to the central dust lane, but the northern plume appears quite real and reaches a projected height of 2.4 $h^{-1}$ kpc from the nucleus. There is another less luminous plume near the northwest nucleus. The primary nucleus accounts for 42% of the total H$\alpha$+[N II] emission, the northwest nucleus contributes 16%, the base of the southern tail contributes another 10%, and the remainder is distributed in knots along the dust lane. There is no ionized gas emission associated with any of the other outer H I features or

UCG 957.

There is a total H I mass of $M_{\rm HI} = 4\times10^9\,h^{-2}\,M_\odot$ in the NGC 520 system, or $4.5\times10^9\,h^{-2}\,M_\odot$ if we correct for the central absorption by interpolating between the adjoining emission. There is $8.8\times10^9\,h^{-2}\,\alpha_{\rm CO}\,M_\odot$ of molecular gas detected out to a radius of 90″ (9.8 $h^{-1}$ kpc) from the primary nucleus (Young *et al.* 1989), making this system even more gas rich for its mass than Arp 295b and NGC 4676a ($M_{gas}/M_{dyn}$=0.14). A third of this gas mass is concentrated in a rotating disk of molecular gas with a radius of 6″ (Sanders *et al.* 1988) roughly centered on the primary nucleus. There is no concentrated CO emission detected at the location of the NW nucleus (Yun & Hibbard, in preparation).

Previous VLA observations of NGC 520 were a factor of four less sensitive ($N_{\rm HI}\gtrsim 8\times10^{19}\,{\rm cm}^{-2}$; Stanford 1990) than those presented here. This led to only a partial mapping of the gas distribution, giving the impression that the H I formed a single tidal feature reaching from the tip of the southern tail and connecting back to UCG 957. The new observations provide a radically different picture. The most striking discovery is the form of the H I velocity field (Fig. 5c & Fig. A-6), which looks remarkably like a single slightly irregular rotating gas disk. We are particularly surprised by the outermost ring of gas which extends not just up to, but smoothly *through* the dwarf galaxy UCG 957. H I is also detected in the dwarf galaxy MCG 00-04-124, which lies 85 $h^{-1}$ kpc to the southeast (Fig. A-6). The rapid drop in the area profile with column density seen in Fig. 9b suggests that we have mapped most of the H I.

There is $3.5\times10^8\,h^{-2}\,M_\odot$ of H I which is unmistakenly associated with the UGC 957, and which has a small velocity off-set (∼10 km s$^{-1}$) from the surrounding gaseous ring (velocity channels from 2127–2191 km s$^{-1}$ in Fig. A-7). From this we deduce that the dwarf is physically associated with the ring, though we cannot tell if it is completely embedded within or lies slightly above or below the ring. The H I column density is highest north and ahead of the dwarf (with respect to the clockwise sense of rotation), while the H I velocity dispersion is enhanced below and behind (Hibbard & van Gorkom 1993). Rudimentary numerical modeling has shown that it is unlikely that the UCG galaxy is responsible for the main optical features of the NGC 520 system (Stanford and Balcells 1991), and it remains a distinct possibility that it has been recently assembled from the surrounding ring of gas (Hibbard & van Gorkom 1993). We note that this system has blue colors (*B-R*=1.0, Hibbard 1995), similar to the southern tail.

Based on the iso-velocity contours (Fig. 5c and Fig. A-6), we separate the H I distribution into four structures: an inner inclined gas disk with the highest column densities ($M_{\rm HI} = 1.6\times10^9\,h^{-2}\,M_\odot$, or $2.2\times10^9\,h^{-2}\,M_\odot$ correcting for absorption); a less inclined intermediate disk ($M_{\rm HI} = 0.6\times10^9\,h^{-2}\,M_\odot$); an incomplete outer gas ring ($M_{\rm HI} = 1.1\times10^9\,h^{-2}\,M_\odot$); and disturbed gas associated with the southern tidal tail which passes through the other three systems and extends to the southeast ($M_{\rm HI} = 0.5\times10^9\,h^{-2}\,M_\odot$), in the direction of MCG+00-04-124. Assuming that the gas is on circular orbits, we fit for rotational velocity, inclination and position angle for separate rings of gas (Begeman 1989). Due to the central absorption and the high inclination of the inner gas disk, we cannot find a solution of the inner 1′, and use the value of 85° given by Bernlöhr (1993). For the intermediate disk we derive $i \approx$ 60°, while for the outer ring we derive $i \approx$ 50°. Surprisingly, there is very little change in the position angle of the line of nodes going from the inner disk to the outer rings, despite the twists in the iso-velocity contours between them.

As discovered by Stanford (1990), the kinematics of the inner H I suggest that it is distributed within a nearly edge-on disk with a radius $\sim$14 $h^{-1}$ kpc, and with its kinematic center near the primary nucleus. What was thought to be a northern tail by Arp (1966) and Toomre (1977) is actually the edge of this disk, which is approaching us. The stellar kinematics of the inner regions show that the NW nucleus is blue-shifted with respect to the primary nucleus (Stanford & Balcells 1990, Bernlöhr 1993) and so the remnant nuclei circle each other in the same direction as the H I disk rotates. There is no H I peak associated with the NW nucleus, although the gas in this region shares in its velocity (2120–2180 $\mathrm{km\,s^{-1}}$, Stanford & Balcells 1990; compare Fig. A-7). The agreement between the stellar velocities of the NW nucleus and the surrounding disk H I confirms that the NW nucleus is embedded within the main body of the merger. The presence of such a regularly rotating disk in the inner regions is rather surprising given the highly disturbed appearance at optical and infrared wavelengths. This observation suggests that *mergers do not necessarily destroy the gas disks of the progenitors.*

The morphology of outer starlight is remarkably different from that of the gas. A bright tail emerges from just south of the remnant, stretching 15 $h^{-1}$ kpc from the main body before bending east and connecting onto a broad low surface brightness plume which extends all the way back to UCG 957 (Fig. 2c; see also Stockton & Bertola 1980). These features are anti-correlated with the outer gas structures. The optical plume is found to be gas-poor along most of its length ($M_{\mathrm{HI}}/L_R < 0.1$ $M_\odot L_\odot^{-1}$), while the outer gaseous structures are largely devoid of light ($M_{\mathrm{HI}}/L_R >$ 10 $M_\odot L_\odot^{-1}$). The peak column densities of the H I in the intermediate disk fall just to the side of the optical plume (Fig. 4c, Fig. A-1c). Similarly, along the southern tail the peaks in the gas lie just to the east of the peak optical surface brightness (Fig. A-1c), giving $M_{\mathrm{HI}}/L_R = 0.2$ $M_\odot L_\odot^{-1}$ on the tail and three times higher behind it. It is interesting that the H$\alpha$ emission lies along the leading edge of this gas, and that the H I velocities deviate most strongly from circular motion (having higher velocities than they would otherwise; Fig. 5c) in this area. The H I and H$\alpha$ kinematics and morphology give the impression that the stellar tail is disturbing the underlying H I distribution from the gas-rich disk it passes by, giving it higher radial velocities and causing it to pool behind the stellar tail in its wake.

Upon first glance it may seem that the southern optical tail gives rise to the outermost gaseous ring, but we do not believe this is the case not only because of the anti-correlation between the outer gaseous and optical features mentioned above, but also because the gas kinematics are dominated by rotation (except, as just mentioned, where the optical tail crosses the H I disks). The restricted azimuthal distribution of the tail-plume system, on the other hand, makes it unlikely that these stars have rotational kinematics. We conclude that the gas rings and optical tail-plume system are kinematically distinct identities. Yet the striking anti-correlation between these features suggest that they somehow "know about" each other. At a rotation velocity of 190 km s$^{-1}$ (Table 3), the observed displacements would be erased in less than a few times $\times 10^7$ years. A similar anti-correlation between outer H I and faint starlight is found in some luminous infrared mergers (Hibbard & Yun in preparation) and in some shell galaxies (Schiminovich *et al.* 1994, 1995).

We suggest that the NGC 520 system is a result of an encounter between a gas-rich system with an extensive disk and a gas poor system, such as an S0 or Sa galaxy. In this interpretation, both the southern optical tail and faint optical plume arose from the gas poor progenitor, and are disturbing the underlying H I distribution from the gas-rich disk as they pass by. The biggest objection to this explanation is the southern-most gaseous feature seen in Fig. 5c (Fig. A-6) which appears to be a natural extension of the southern tail, and would therefore seem to indicate two gas rich progenitors. It is possible that this extension is left over from a previous passage, or is a continuation of the outermost ring, or was perhaps torn off of MCG+00-04-124, but the issue is far from resolved.

### A.4 NGC 3921

NGC 3921 (=Arp 224=Mkn 430) was recently the focus of a detailed study by Schweizer (1995), who found compelling evidence for a merger origin. In addition to the two tidal extensions which warranted its inclusion in Toomre's sequence, this system exhibits a single nucleus with a centrally relaxed light profile in the $g,V$ and $K$-bands (Schweizer 1995, Stanford & Bushouse 1991), and obeys the Faber-Jackson relationship for ellipticals (Lake & Dressler 1986). There are many examples of loops, ripples, and other fine-structure surrounding the remnant (Schweizer & Seitzer 1992, Schweizer 1995), and the optical isophotes are off-centered and "sloshing" (Schweizer 1995), indicating that the luminous matter is not yet in equilibrium and that the remnant is dynamically young. The main body exhibits the strong Balmer absorption lines typical of a post-starburst population (Schweizer 1990, 1995; Kennicutt 1992b). These line strengths, together with the luminosity, $UBV$ colors, color gradients, velocity dispersion, and fine-structure index suggest that NGC 3921 is a 0.7±0.3 Gyr old merger remnant well on its way to becoming an elliptical in its photometric properties (Schweizer 1995).

This system has two companions detected in the H I observation: NGC 3916, an Sb spiral with a warped disk containing $2 \times 10^9 \, h^{-2} \, M_\odot$ of H I located 80 $h^{-1}$ kpc to the northwest, and MCG +09-19-213 (=Sey 131), a dwarf galaxy located 85 $h^{-1}$ kpc to the west. Ionized gas measurements reveal that the compact system located 65″ west and 30″ south of NGC 3921 has a recessional velocity of 5712 km s$^{-1}$ (Schweizer 1995), and is therefore a nearby companion ($\Delta r$=20 $h^{-1}$ kpc). This system is seen on the edge of Fig. A-1d, and is discussed below. The system which appears to the NE in the deep $R$-band image (Fig. 2d) is the brightest member of the background cluster Abell 1400 (Schweizer 1995).

The H II emission is primarily associated with the main body (95%), and very centrally concentrated (see also Schweizer 1995): more than half is emitted from the central 1 $h^{-1}$ kpc (FWHM). In contrast to the H$\alpha$ morphology of the earlier-stage systems, the central emission here is rather regular (Fig. 6e).

There are additional ionized gas emission features to the west and southwest of this system (Fig. 6e). The southwestern feature lies along an enhancement in the H I column density, while the western feature has no associated H I emission (Fig. A-1d). These systems are symmetrically placed with respect to the compact galaxy mentioned above and seen to the far right of Fig. 6e, and the H I associated with the southern H$\alpha$ emission feature borders the companion over the velocity range 5795–5849 km s$^{-1}$ (Fig. A-9). While these velocities are 80–140 km s$^{-1}$ higher than the recessional velocity of the companion, they do appear somehow associated, although their origin and nature remain enigmatic.

There are similar quantities of molecular and atomic hydrogen in NGC 3921. All of the molecular gas resides within the main body ($r <7$ $h^{-1}$ kpc; Combes & Casoli, in preparation). More than half of this is localized within a small knot with dimensions 7″×4″ (2 $h^{-1}$ kpc ×1 $h^{-1}$ kpc) (Yun & Hibbard, in preparation), which is presumably associated with the nuclear H$\alpha$ emission. In contrast, the majority of the atomic gas is found outside of the main body. The atomic hydrogen which does lie within the main body is spread over a 270 km s$^{-1}$ velocity range roughly centered on the systemic velocity. This gas has very disordered kinematics (Fig. 5d; the central regions in this figure are colored green), with blue-shifted gas to the north and south and red-shifted gas to the west. This is similar to what is found from the kinematics of the central ionized gas (Schweizer 1995).

The northern plume is quite bright ($\mu_R \sim 23.5$ mag arcsec$^{-2}$) and contains most of the outer luminosity, but no gas ($M_{\rm HI}/L_R < 0.02$ $M_\odot L_\odot^{-1}$). It appears to be a continuation of the luminous southern loop system which is also gas poor (Fig. A-1d). For the purposes of the Table 5 entry we measure the length of this feature from the nucleus along the loop to the end of the plume. Fig. A-1d shows that the atomic hydrogen in the vicinity of this loop is displaced to the outside of the starlight, and the H I kinematics suggest that it is associated with the gas in the southern tail (Fig. A-9), rather than another tidal feature. That this entire loop-plume structure is atomic gas poor may indicate that the progenitor was gas poor. The striking anti-correlation between the starlight and gas seen in Fig. A-1d suggests that, as was the case for the stellar plume and outer H I in NGC 520, these separate components somehow "know about" each other.

The base of the southern tail is bright, but due to projection effects it is impos-sible to tell how it connects back to the remnant body. Beyond the bright southern loop, the southern tail is considerably fainter ($\mu_R \gtrsim 25$ mag arcsec$^{-2}$), and contains only a small fraction of the total luminosity ($3\times10^8\,h^{-2}\,L_\odot$). However this structure contains most of the outer H I ($\sim 2\times10^9\,h^{-2}\,M_\odot$), and becomes progressively more gas rich along its length (Fig. 7b). The H I tail does not end with the optical tail, but bends back around in a continuous structure to the west and then north again. We believe that the tail does not actually approach the main body again, but just appears to in projection.

Outside the main body, the southern tail shows a smooth and continuous velocity structure (Figs. A-8 and 5d). At the base of the H I tail, just south of the nucleus, the gas has a velocity of -30 km s$^{-1}$, reaches a minimum of -120 at the point of peak H I emission in Fig. A-1d (blue in Fig. 5d), and thereafter increases smoothly. Further along the tail the velocities cross through systemic (yellow), become red-shifted by +30 km s$^{-1}$ at the H II region at the end of the optical tail, and reach an maximum of +70 km s$^{-1}$ at the southwestern-most tip.

To constrain the gas motions in the tail, we need to know whether the southern tail passes in front of or behind the remnant. This cannot be determined with any degree of confidence (see also Schweizer 1995). The lack of a significant dust feature associated with the gas-rich southern tail may suggest that the tail lies behind the remnant and southern loop system, but it is possible that the tail contains very little dust. If the tail lies behind the remnant, then the velocity field indicates that the blue-shifted material at the base of the tail is falling back toward the remnant, while the southern-most regions are still expanding. In this case the tail is swinging away from us, with its angular momentum vector pointing to the west. If instead the tail crosses the face of the remnant, the velocity field indicates that the tail swings towards us, with its angular momentum vector pointing to the east. The velocity reversals near the end of the optical tail indicate that at this point the tail bends away from us.

The southern tail contains a ridge of star forming regions along its base, accounting for 3% of the H$\alpha$ emission. Beyond this base, the optical tail appears as a collection of about five knots connected by fainter light. At the end of the optical tail there is a concentration of atomic gas, starlight, and H$\alpha$ emission ($M_{\rm HI}$ $\sim 10^8\,h^{-2}\,M_\odot$, $L_R$ $\sim 3\times10^7\,h^{-2}\,L_\odot$, $L_{H\alpha+[NII]}$ $\sim 8\times10^{37}\,h^{-2}\,{\rm erg\,s}^{-1}$). We find that the atomic hydrogen line-widths are increased at this concentration and we suggest that it might be self-gravitating, in which case it has observational characteristics typical of dwarf galaxies (Hibbard & van Gorkom 1993).

TABLE A-1. Properties of HI Detected Systems.

| NAME | $\alpha_{1950}{}^{(1)}$ | $\delta_{1950}{}^{(1)}$ | $M_{\rm HI}$ $h^{-2}{\rm M}_\odot$ | $V_o$ km s$^{-1}$ | $\Delta V^{(2)}$ km s$^{-1}$ |
|---|---|---|---|---|---|
| **Arp 295 System** | | | | | |
| G1 | $23^h38^m38.8^s$ | $-03°54'31''$ | $6.0\times10^8$ | 6840 | 6790–6890 |
| G2 | $23^h39^m04.0^s$ | $-03°55'58''$ | $3.0\times10^7$ | 6740 | 6710–6750 |
| G3 | $23^h39^m06.7^s$ | $-03°56'03''$ | $4.7\times10^8$ | 6590 | 6570–6610 |
| G4 | $23^h39^m09.4^s$ | $-03°56'04''$ | $1.6\times10^9$ | 6610 | 6650–6570 |
| G5 | $23^h39^m13.1^s$ | $-03°46'25''$ | $1.2\times10^9$ | 6905 | 6890–6910 |
| A295a | $23^h39^m13.1^s$ | $-03°56'41''$ | $2.4\times10^9$ | 6800 | 6530–7050 |
| A295b | $23^h39^m26.6^s$ | $-03°53'32''$ | $8.3\times10^9$ | 6950 | 6810–7090 |
| G6 | $23^h39^m30.2^s$ | $-03°51'30''$ | $9.5\times10^8$ | 6780 | 6830–6710 |
| G7 | $23^h39^m34.1^s$ | $-03°46'41''$ | $2.0\times10^8$ | 6850 | 6790–6910 |
| G8 | $23^h39^m35.3^s$ | $-03°48'17''$ | $1.0\times10^8$ | 6820 | 6800–6840 |
| G9 | $23^h40^m27.2^s$ | $-04°07'57''$ | $5.0\times10^8$ | 6800 | 6790–6830 |
| **NGC 4676 System** | | | | | |
| N4676a | $12^h43^m44.1^s$ | $+31°00'14''$ | $1.8\times10^9$ | 6680 | 6530–6930 |
| N4676b | $12^h43^m45.3^s$ | $+30°59'43''$ | $2.0\times10^9$ | 6510 | 6250–6750 |
| S comp | $12^h44^m01.1^s$ | $+30°47'17''$ | $1.2\times10^8$ | 6600 | 6530–6670 |
| **NGC 520 System** | | | | | |
| N520 | $01^h21^m59.6^s$ | $+03°31'53''$ | $>3.9\times10^{9(3)}$ | 2260 | 2050–2500 |
| U957 | $01^h21^m49.1^s$ | $+03°37'20''$ | $3.5\times10^8$ | 2150 | 2100–2190 |
| MCG +00-04-124 | $01^h22^m33.4^s$ | $+03°21'16''$ | $5.3\times10^8$ | 2240 | 2150–2310 |
| **NGC 3921 System** | | | | | |
| N3921 | $11^h48^m28.9^s$ | $+55°21'28''$ | $2.9\times10^9$ | 5880 | 5720–5990 |
| N3916 | $11^h48^m12.7^s$ | $+55°25'17''$ | $2.1\times10^9$ | 5810 | 5590–6010 |
| MCG +09-19-213$^{(4)}$ | $11^h47^m54.4^s$ | $+55°20'10''$ | $>4\times10^8$ | 5620? | 5610–>5680 |
| **NGC 7252 System** | | | | | |
| N7252 | $22^h17^m57.9^s$ | $-24°55'50''$ | $2.0\times10^9$ | 4740 | 4520–4880 |
| NE comp | $22^h18^m33.2^s$ | $-24°46'35''$ | $7.6\times10^8$ | 4795 | 4760–4825 |

Notes to TABLE A-1

(1) Positions as measured from a local continuum peak, if any, or from MANN measuring engine measurements. These positions are probably accurate to no better than a few arcsecs.

(2) $\Delta V$ represent the range of channels showing H I emission. It is often larger than the velocity extremes given in the first-moment maps, since the latter represent intensity weighted velocity centroids.

(3) Due to absorption against a central continuum source, the H I mass in NGC 520 is a lower limit.

(4) The H I emission from MCG +09-19-213 appeared at the edge of the passband, and is not completely measured in these observations. The H I mass is therefore a lower limit, and the upper velocity limit unknown.

**FIGURE AI-1:** Details on differences between the H I to optical distribution along the sequence. In each of the panels, the deep $R$-band observations are shown in negative greyscale, the H I emission is in black or white contours, and regions with detectable H$\alpha$+[N II] emission are white. In the top two panels the natural weighted H I data has been used for maximum sensitivity, while for the bottom two the uniform weighted data has been used for maximum spatial resolution **(a)** The southern tail of Arp 295a, showing the displacement between the gas and light peaks along the tail, and that the optical tail extends beyond the end of the gas distribution. **(b)** The northern tail of NGC 4676a, showing that again the optical light extends well beyond the end of the gas emission. **(c)** NGC 520, showing that the H I column density peaks lie *behind* the bright southern optical tail, and that the outer gas to the north clearly lies just off the edge of the optical plume (top middle of frame). Crosses mark the location of the primary and secondary nucleus. **(d)** NGC 3921, showing that the southern loop system is gas poor except for where the gas from the southern tail is projected across the it, and that the peaks in the H I are displaced to the edge of the southern loop feature.

**FIGURE A-2:** Contours of the H I intensity weighted velocity (moment 1) drawn on a negative greyscale of the integrated H I emission (moment 0) of the Arp 295 system. The contour spacing is 20 km s$^{-1}$, and the greyscale range is from $10^{19}$ cm$^{-2}$ (white) to $2\times10^{21}$ cm$^{-2}$ (black). The natural weighted C+D array data has been used. Companions G1–8 are labeled with their systemic velocities. The last companion (G9) lies outside the scope of this map, about 21′ (420 $h^{-1}$ kpc) to the southeast. Indicated to either side of most systems are their peak velocities, relative to the local systemic (*e.g.*, the ±50 labeled to the north and south of G1 indicate velocities of 6840±50 km s$^{-1}$ for this system). The velocities along the bridge are given relative to Arp 295b, and the velocities along the southern tail are given relative to Arp 295a.

**FIGURE A-3:** Channel maps of the H I emission in the Arp 295 system. The natural weighted C+D array VLA data is shown at a channel separation of 21.6 km s$^{-1}$. The HPBW of the synthesized beam (26″) is indicated in the lower left hand corner, and crosses mark the location of Arp 295a&b and two companions (G4 and G6 in Table A-1). The heliocentric velocities (km s$^{-1}$) are given in the upper left hand corner of each frame. The contour levels are (-4,4,8,12,25,50) ×10$^{19}$ cm$^{-2}$, where 10$^{19}$ cm$^{-2}$ = 0.28 mJy beam$^{-1}$, and the lowest contour corresponds to 5 times the rms noise.

**Figure A-3**, continued.

**FIGURE A-4:** Contours of the the H I intensity weighted velocity field (moment 1) drawn on a negative greyscale of the integrated H I emission (moment 0) of NGC 4676. The contour spacing is 20 $\mathrm{km\,s^{-1}}$, and the greyscale range is from $10^{19}\,\mathrm{cm^{-2}}$ (white) to $2\times10^{21}\,\mathrm{cm^{-2}}$ (black). The natural weighted C+D array data has been used. The systemic velocities of both NGC 4676a&b are listed below their labels, and tail velocities relative to the total systemic (6600 $\mathrm{km\,s^{-1}}$) are listed to the side.

**FIGURE A-5:** Channel maps of the H I emission in NGC 4676. The natural weighted C+D array VLA data is shown at a channel spacing of 43 km s$^{-1}$, with the heliocentric velocities (km s$^{-1}$) given in the upper left hand corner of each frame. The HPBW of the synthesized beam (20″) is indicated in the lower left hand corner, and crosses mark the location of NGC 4676a&b. The contour levels are -5,5,10,15,20,30,50) ×10$^{19}$ cm$^{-2}$, where 10$^{19}$ cm$^{-2}$ = 0.08 mJy beam$^{-1}$, and the lowest contour corresponds to 2.5 times the rms noise.

**Figure A-5, continued.**

**FIGURE A-6:** Contours of the the H I intensity weighted velocity field (moment 1) drawn on a negative greyscale of the integrated H I emission (moment 0) of the NGC 520 system. The contour spacing is 20 km s$^{-1}$, and the greyscale range is from $10^{19}$ cm$^{-2}$ (white) to $2\times10^{21}$ cm$^{-2}$ (black). The natural weighted C+D array data has been used. The systemic velocities of all systems are listed below their labels, and velocities relative to either NGC 520 or MCG+00-040124 are listed to the side. Absorption against a central extended radio source responsible for the central "hole". The three-quarters complete outer ring passing through UGC 957 is very prominent in this figure, and the isovelocity contours of the entire system exhibit the typical "spider diagram" of a rotating disk, albeit with some peculiarities. Note that the position angle of the maximum velocities of the inner disk is approximately the same as in the outer ring. Also note the southern tidal extension that points roughly towards MCG+00-04-124.

**FIGURE A-7:** Channel maps of the H I emission in NGC 520. The natural weighted C+D array VLA data is shown after smoothing to a channel separation of 21 km s$^{-1}$, with the heliocentric velocities (km s$^{-1}$) given in the upper left hand corner of each frame. The HPBW of the synthesized beam (24″) is indicated in the lower left hand corner, and crosses mark the location of both the primary and secondary nuclei of NGC 520, and of UCG 957, MCG+00-04-124, and two stars. The contour levels are (-2,2,4,8,16,32) ×10$^{19}$ cm$^{-2}$, where 10$^{19}$ cm$^{-2}$ = 0.54 mJy beam$^{-1}$, and the lowest contour corresponds to 4 times the rms noise.

**Figure A-7, continued.**

**FIGURE A-8:** Contours of the the H I intensity weighted velocity field (moment 1) drawn on a negative greyscale of the integrated H I emission (moment 0) of the NGC 3921 system. The contour spacing is 20 km s$^{-1}$, and the greyscale range is from $10^{19}$ cm$^{-2}$ (white) to $2\times10^{21}$ cm$^{-2}$ (black). The natural weighted C+D array data has been used. The systemic velocities of all systems are listed below their labels, and velocities relative to the local systemic are listed to the side. The H I emission associated with MCG +09-19-213 has not been fully mapped, as this system appeared at the edge of the observed passband, with the negative velocities missing.

**FIGURE A-9:** Channel maps of the H I emission in NGC 3921. The natural weighted C+D array VLA data is shown at a channel separation of 10.7 km s$^{-1}$, with the heliocentric velocities (km s$^{-1}$) given in the upper left hand corner of each frame. The HPBW of the synthesized beam (23″) is indicated in the lower left hand corner, and a cross marks the location of the NGC 3921 nucleus. The contour levels are (-2,2,4,8,16,32,64) $\times 10^{19}$ cm$^{-2}$, where $10^{19}$ cm$^{-2}$ = 0.44 mJy beam$^{-1}$, and the lowest contour corresponds to 4 times the rms noise.

**Figure A-9, continued.**